\documentclass{SCGE}

\allowdisplaybreaks
\usepackage{graphicx}
\usepackage{amssymb}
\usepackage{mathrsfs}
\usepackage{epstopdf}
\usepackage{ulem}
\usepackage{subfigure}
\usepackage{float}

\newcommand\bi{\bibitem}
\newcommand\be{\begin{eqnarray}}
\newcommand\ee{\end{eqnarray}}
\newcommand\la{\langle}
\newcommand\ra{\rangle}
\def\roughly#1{\mathrel{\raise.3ex\hbox{$#1$\kern-.75em%
\lower1ex\hbox{$\sim$}}}}
\def\lsim{\roughly<}
\def\gsim{\roughly>}
\begin{document}

%%%%%%
\begin{picture}(0,0){\rm
\put(0,-20){\makebox[160truemm][l]{\bf {\sanhao\raisebox{2pt}{.}}
Invited Review  {\sanhao\raisebox{1.5pt}{.}}}}}
%\put(0,-34){\jiuwuhao {\textcolor[rgb]{0.5,0.5,0.5}{\sf Progress of Projects Supported by NSFC
%}}}
\end{picture}

\def\bm{\boldsymbol}

\def\du{\end{document}}

% The author doesn't need fill in it.
\Year{2016} %
\Month{Month} %
\Vol{***} %  卷号
\No{***} %  期号
\BeginPage{1} % 起页码
\AuthorMark{{\rm Yong-Liang Ma}, et al.}  %(11月注释：页眉上的作者)
\AuthorMarkCite{{\rm Yong-Liang Ma, Mannque Rho}. } %(11月注释：citation中的作者)
\DOI{********} % The author doesn't need fill in it.
\ArtNo{******}

\title{Recent progress on dense nuclear matter in skyrmion approaches}

\author[1]{Yong-Liang Ma}{}
%\email[1]{yongliangma@jlu.edu.cn}
\author[2]{Mannque Rho}{}
%\email[2]{mannque.rho@cea.fr}

\address[{\rm1}]{Center of Theoretical Physics and College of Physics, Jilin University, Changchun, 130012, China;\\
Email:yongliangma@jlu.edu.cn }
\address[{\rm2}]{Institut de Physique Th\'eorique, CEA Saclay, 91191 Gif-sur-Yvette c\'edex, France;\\
Email:mannque.rho@cea.fr}

\maketitle
%\vspace{-3.5mm}{
%\footnotesize\begin{center} Received October 26, 2014; accepted November 5, 2014
%\end{center}
%}\vspace*{-5mm}

% Abstract is required.
\begin{center}
\rule{16.5cm}{0.4pt}
\parbox{16.5cm}
{\begin{abstract}
The Skyrme model provides a novel unified approach to nuclear physics. In this approach, single baryon, baryonic matter and medium-modified hadron properties are treated on the same footing. Intrinsic density dependence (IDD) reflecting the change of vacuum by compressed baryonic matter figures naturally in the approach. In this article, we review the recent progress on accessing dense nuclear matter by putting baryons treated as solitons, namely, skyrmions, on crystal lattice with accents on the implications in compact stars.
\end{abstract}}
\end{center}

\vspace*{-0.6cm}

\begin{center}
\parbox{16.5cm}
{\bf\jiuhao Skyrme model, hidden symmetries, skyrmion crystal, holographic nucleon, BPS nuclei, compact stars.}%关键词
\end{center}

\begin{center}
% 12.39.Dc Skyrmions
% 12.39.Fe Chiral Lagrangians
% 14.20.Dh Protons and neutrons
% 21.65.-f Nuclear matter
{\PACS{\rm 12.39.Dc, 21.65.-f, 12.39.Fe}}%分类号
\CITA    %%(11月注释：Citation内容自动生成)
%\Cit{~~~???, et al. ???. Sci China-Phys Mech Astron, 2014, 57: 1--6, doi:}%%(11月注释：Citation内容需手动填写)
\end{center}

\textwidth=178truemm \textheight=236truemm%%%%%%新版式要加上

%%%%%%%%%%%%%%%%%%%%%%%%%%%%%%%%%%%%%%%%%%%%%%%%%%%%%%%%%%%%
\wuhao\vspace*{1.5mm}

\begin{multicols}{2}

\tableofcontents

%%%%%%%%%%%%%%%%%%%%%%%%%%%%%%%%%%%%%%%%%%%%%%%%%%%%%%%%%%%%
%% Text of article.
%%%%%%%%%%%%%%%%%%%%%%%%%%%%%%%%%%%%%%%%%%%%%%%%%%%%%%%%%%%%
%    Section headings
\renewcommand{\baselinestretch}{1.08} \baselineskip 12.2pt\parindent=10.8pt

\renewcommand{\thefootnote}{\#\arabic{footnote}}

\section{Introduction}

Accessing the properties of single nucleon and multi-nucleon matter (or nuclear matter) in terms of skyrmions~\cite{Skyrme:1961vq}~\footnote{By ``skyrmion" will be meant the generic soliton model for baryons without specification of the degrees of freedom involved in the Lagrangian.  The soliton model built on the Skyrme Lagrangian consisting of the quadratic current-algebra term and the Skyrme quartic term {\it only} will be referred to as ``Skyrme model."  Whatever the degrees of freedom involved, the topology is carried by the pion field only.} has become an exciting new topic in nuclear physics on par with condensed matter physics and string theory~\cite{BReditor}.  In the limit of  large number of colors $N_c$~\cite{'tHooft:1973jz}, the skyrmion is a baryon in QCD and shares the same properties as constituent quark model ~\cite{Witten:1979kh,manohar-largeN}. From then on, Skyrme's idea has been widely accepted in the physics community and indeed, as one can see in \cite{BReditor}, has been spectacularly successful in condensed matter physics and is beginning to have significant impact in nuclear physics. Earlier achievements were reviewed in Refs.~\cite{Zahed:1986qz,Holzwarth:1985rb}.

Since the 1980s, especially in the past decade,  a considerable progress has been made on describing dense nuclear matter by putting skyrmions on crystal lattice, the strategy initiated by Klebanov~\cite{Klebanov:1985qi}.  In this review we summarize what we have learned from that approach on compressed baryonic matter, an area of nuclear dynamics for which there are no well-established theoretical tools or experimental data. The lattice QCD technique, fruitfully applied to high temperature matter, is presently powerless in addressing dense matter and effective field theory models, presently available, or phenomenological models based on energy-density functionals are not predictive enough beyond the density regime near normal nuclear matter density.  It will be seen below that the regime of large density above that of nuclear matter $n_0\simeq 0.16$ fm$^{-3}$ is a total wilderness.

In the skyrmion approach, one starts with an effective Lagrangian, with as many relevant degrees of freedom as relevant taken into account, and puts the resulting skyrmions on the crystal lattice.   It has become clear from the modern development of holographic QCD for baryons (see \cite{BReditor} for review) that, while the pion is essential for supplying topology, there is no reason to suggest that the pion is enough to arrive at a realistic description of the baryon and in particular baryonic matter. In fact an infinite tower of vector mesons do and must intervene. Given a realistic Lagrangian for skyrmion structure,  dense matter can then be simulated by putting multi-skyrmions on crystal lattice and squeezing the crystal size. The Skyrme model can therefore provide us a unified framework to study the single baryon, baryonic matter and medium modified hadron properties~\cite{Lee:2003aq}. We will briefly discuss the basics of nuclear matter from skyrmions and the potential power of the skyrmion model over other nuclear effective field theories in Sec.~\ref{sec:ABC} and Sec.~\ref{sec:power}.

Since the discovery of the AdS/CFT correspondence~\cite{Maldacena:1997re}, the holographic QCD model~\cite{DaRold:2005zs,Erlich:2005qh,Sakai:2004cn,Sakai:2005yt,Hong:2007kx,Hong:2007ay} has provided a novel approach to the strong interaction processes at low energy. In this approach, one can access not only the elementary meson and baryon dynamics~\cite{DaRold:2005zs,Erlich:2005qh,Sakai:2004cn,Sakai:2005yt,Hong:2007kx,Hong:2007ay} but more significantly the dense baryonic matter using the topology profiles that is difficult to access by methods that do not resort to topology~\cite{Rho:2009ym,Gorsky:2013dda,Bolognesi:2013nja,Kaplunovsky:2012gb,Elliot-Ripley:2016uwb,Elliot-Ripley:2016ctk}. We will review the progress in this direction in Sec.~\ref{sec:holography}. Special emphasis is put  on the impact  of the infinite tower of vector mesons on multi-skyrmion properties.

We have learned from the long history of nuclear phenomenology that not only the pseudoscalar mesons, pions, but also the lowest-lying vector mesons, $V=(\rho, \omega)$  and the scalar meson (that we shall call $\sigma$)  of comparable mass are crucial for understanding the nuclear forces. In Sec.~\ref{sec:vector}, we review the effect of the lowest-lying vector and scalar mesons on nucleon and nuclear matter properties by using a ``chiral-scalar effective theory" implementing hidden local symmetry (HLS)~\cite{Bando:1984ej,Harada:2003jx} and hidden scale symmetry.

In Sec.~\ref{sec:BPS}, we discuss a variant soliton model which focuses on the BPS structure of heavy nuclei~\cite{Adam:2010fg,Adam:2010ds}. This model departs from the standard approach to skyrmion which is anchored on chiral Lagrangians that describe low-energy nonperturbative strong interactions typified by the nonlinear sigma model. It starts from the characteristic features of heavy nuclei and nuclear matter of incompressible liquid structure and aims, ultimately, to go to the regime where soft-pion dynamics are recovered so as to make a contact the premise of QCD. In this model, the empirically small value of the nuclear binding energy, especially for nuclei with large mass number $A$, can be well reproduced~\cite{Adam:2013tda,Adam:2013wya}. In addition, it has been suggested that the BPS model can also give predictions of the mass and radius of massive compact stars~\cite{Adam:2014dqa},  addressing the recent observations~\cite{Demorest:2010bx,Antoniadis:2013pzd}.

The HLS\footnote{Unless specified otherwise, by HLS, we will mean hidden local symmetry with the lowest-lying vector mesons $V$. }, constructed  up to, and including, $O(p^4)$ terms used in the simulation of skyrmion matter,  is quite involved with 17 terms. It turns out however that one can simplify considerably the full HLS Lagrangian by numerically analyzing the contributions from each term of the Lagrangian to the skyrmion and skyrmion matter. In Sec.~\ref{subsec:SimplifyHLS}, it is shown that one can reduce the 17 terms down to 7 terms with an accuracy of $\sim 97\%$.

Finally translating the features obtained in the skyrmion crystal simulation, with focus on the special role of the topology change in the nuclear tensor force, into the ``bare parameters" of the effective baryonic Lagrangian that incorporates hidden symmetries -- hidden local symmetry and scale symmetry,  the properties of both nuclear and dense matter could be addressed in a unified way. This development, not widely appreciated in nuclear physics community, is briefly discussed in Sec.~\ref{sec:matter}.

We finally provide our perspectives and discussions in Sec.~\ref{sec:dis}.

\section{ABC of Baryonic Matter from Skyrmions}

\label{sec:ABC}

The original Skyrme model is a model describing baryons in a mesonic theory in which the chiral symmetry is realized nonlinearly with pion field only~\cite{Skyrme:1961vq}.

In the nonlinear realization of chiral symmetry, the pion figures as the Nambu-Goldstone (NG for short) boson of the spontaneous breaking of chiral symmetry. It is expressed in the polar parameterization through
\begin{eqnarray}
U(x) & = & \exp\left(2i\pi^a T^a/f_\pi\right),
\label{eq:defU}
\end{eqnarray}
where $T^a$ is the generator of the $SU(2)$ group satisfying ${\rm Tr}(T^a T^b) = \frac{1}{2}\delta^{ab}$ and $f_\pi$ is the pion decay constant with an empirical value $f_\pi \simeq 92.4$~MeV.

Since the unitary field $U(x)$ satisfies $U(x)U(x)^\dagger = U(x)^\dagger U(x) = 1$, for any fixed time, say,  $t_0$, the matrix $U(\mathbf{x},t_0)$ defines a map from $R^3$ to the manifold $S^3$  in isospin space. At the low-energy limit, QCD goes to the vacuum, i.e.,
\begin{eqnarray}
U(|\mathbf{x}| \to \infty, t_0)& = & \mathbf{1},
\label{eq:configuxinfinite}
\end{eqnarray}
therefore, all the points at $|\mathbf{x}| \to \infty$ are mapped onto the north pole of $S^3$ and energy of the system is finite. We have the nontrivial map
\begin{eqnarray}
U(\mathbf{x}, t_0): R^3 \to S^3, \label{eq:mapU2S3}
\end{eqnarray}
for the static configuration $U(\mathbf{x}, t_0)$. In the language of topology, these maps constitute the third homotopy group $\pi_3(S^{\,3}) \sim Z$ with $Z$ being the additive group of integers which accounts for the times that $S^3$ is covered by the mapping $U(\mathbf{x}, t_0)$, i.e., winding numbers. Because a change of the time coordinate can be regarded as a homotopy transformation which cannot transit between the field configuration in homotopically distinct classes, the winding number is a conserved quantity in the homotopy transformation by the unitary condition of the field $U(x)$ and condition \eqref{eq:configuxinfinite}. In skyrmion models, the conserved winding number represents the conserved baryon number in QCD. The baryon arises as a topological soliton with the topology lodged in the chiral field $U(x)$.

Note that, in the construction of the skyrmion-type model, the unitary condition of the field $U(x)$ and the condition \eqref{eq:configuxinfinite}  are the essential characteristics that should be taken into account. Indeed, in the construction of the BPS-type skyrmion proposed by Adam {\it et. al.}~\cite{Adam:2010fg,Adam:2010ds} that will be discussed later, only these two conditions are imposed.

\begin{figure}[H]\centering
\includegraphics[scale=0.38]{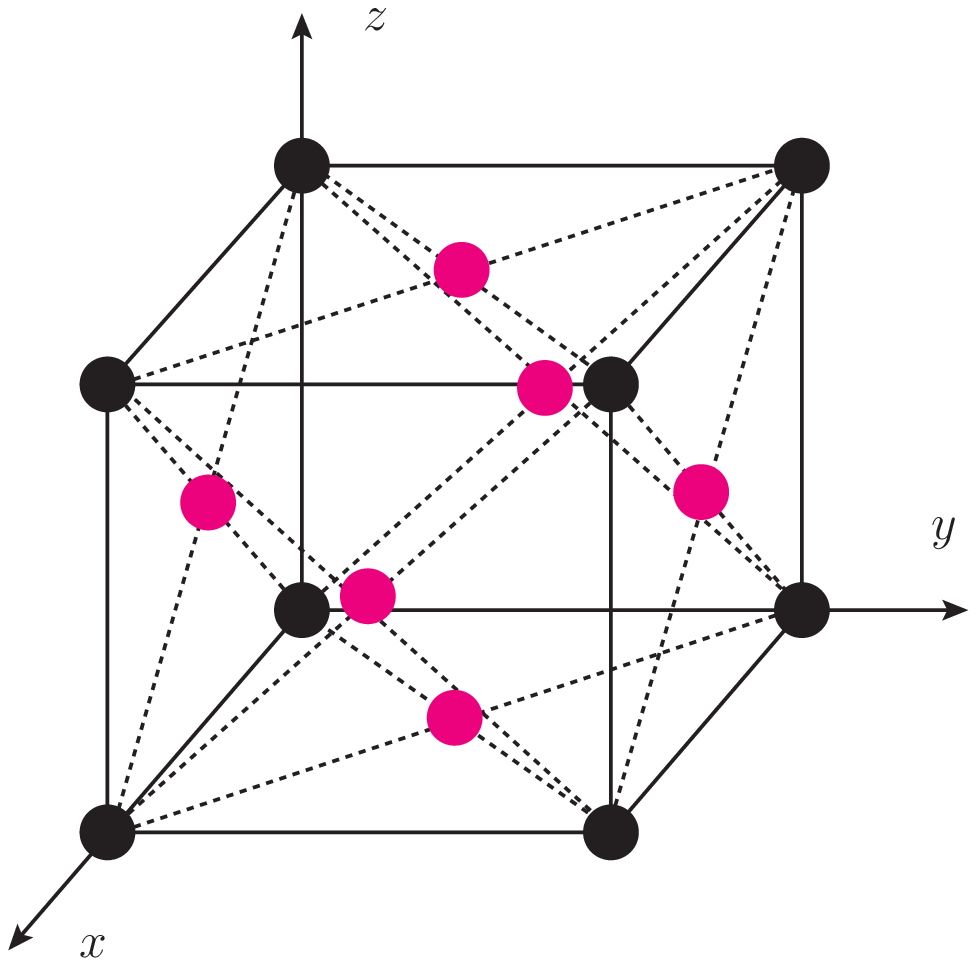}\includegraphics[scale=0.17]{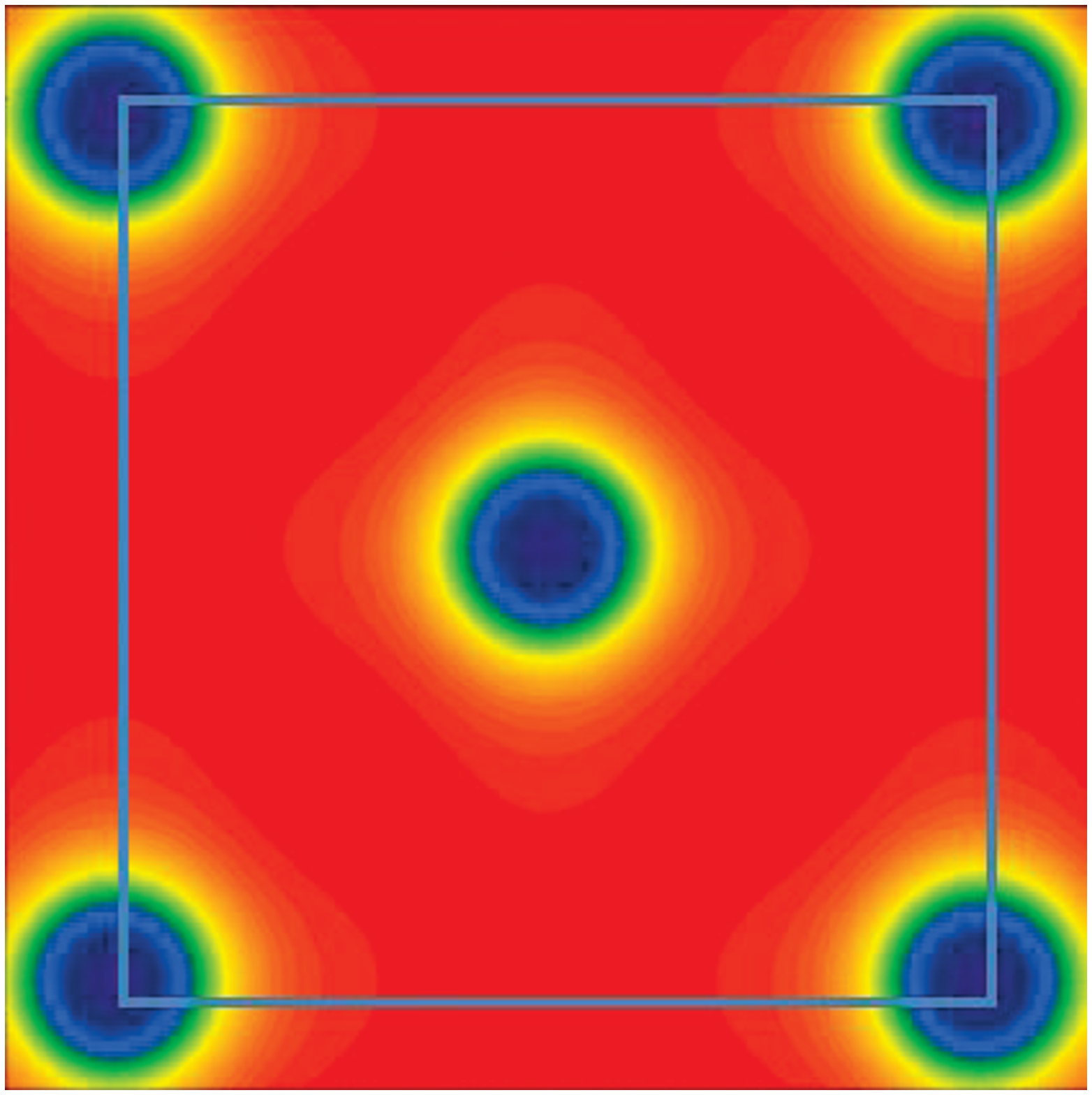}
\caption{The arrangement of skyrmions on the FCC crystal lattice {(left panel)} and baryon density distribution {(right panel)}. {In the left panel, the black dots stand for skyrmions at the lattice vertices and the red dots represent skyrmions on the face centers.}}
\label{fig:fccdis}
\end{figure}

In the skyrmion approach to baryon-baryon interactions, the baryon-baryon interaction at large separation is approximated by the product ansatz~\cite{Skyrme:1961vq} of  two undistorted skyrmions with a relative rotation in spin-isospin space
\begin{eqnarray}
U_{00}(\mathbf{x},\mathbf{x}_1,\mathbf{x}_2) & = &
U_{0}(\mathbf{x}-\mathbf{x}_1)A(\bm{\alpha})U_{0}(\mathbf{x}-\mathbf{x}_2)A^\dag(\bm{\alpha}),\label{eq:productansatzskyr}
\end{eqnarray}
where $A(\bm{\alpha}) = \exp(i\bm{\tau}\cdot\bm{\alpha}/2)$ is the rotation in the isospin space with rotation angle $\alpha$, $U_0(\mathbf{x})$ is the skyrmion configuration satisfying the classical equation of motion of soliton, $\mathbf{x}_1$ and $\mathbf{x}_2$ are the centers of the two skyrmions and $r = |\mathbf{x}_2 - \mathbf{x}_1|$ is the distance. To get the most attractive potential, the pair of skyrmions should be arranged in such a way that they should mutually rotate in the isospin space by angle $\pi$ about the axis perpendicular to the line joining them. One can simulate the nuclear matter by putting skyrmions onto the crystal lattice, first put forward by Klebanov~\cite{Klebanov:1985qi}, and regarding the skyrmion matter as baryonic matter. The density effect enters when the crystal size is changed. Subsequently, Kugler and Shtrikman proposed a new crystal structure~\cite{Kugler:1988mu,Kugler:1989uc}, the face-centered-cubic (FCC) crystal, which gave the minimal energy of the nuclear matter about $4\%$ lower than that simulated by the cubic crystal used by Klebanov. The arrangement of  skyrmions on an FCC crystal and the baryon number density distribution in the crystal face are shown in Fig.~\ref{fig:fccdis}. So far the FCC crystal is the lowest energy crystal configuration known for nuclear ground state.

In the FCC configuration, the skyrmion number density, which is the baryon number density with the skyrmion matter regarded as baryon matter, is expressed as $1/(2L^3)$ where  $2L$  is the crystal size. The size of the crystal lattice corresponding to the normal nuclear density $n_0 = 0.17~$fm$^{-3}$ is then given by $L = 1.43~$fm.

To simulate the nuclear matter in the skyrmion crystal approach, it is convenient to decompose the skyrmion soliton configuration $U_0$ as
 \begin{eqnarray}
 U_0 & = & \phi_0+i \vec{\tau}\cdot\vec{\phi}, \label{eq:definenorphi}
\end{eqnarray}
with the constraint $\phi_0^2 + \vec{\phi}^2 = 1$. In this decomposition, $\phi_0$ accounts for the magnitude of the quark-antiquark condensate $\langle\bar{q}q\rangle$ which in the matter-free space satisfies $\phi_0(\bm{x} \to \infty) \to 1$ and $\phi_a(\bm{x}) \to 0,(a=1,2,3),$ due to the parity invariance of the strong interactions. However, in the inhomogeneous medium, due to the interaction of the pion fluctuation with the medium, the quantities $\phi_0(\bm{x})$ and $\phi_a(\bm{x})$ measure the inhomogeneous quark condensate in the medium~\cite{Harada:2015lma}.

One convenient way to simulate the skyrmion matter in the crystal is to introduce the unnormalized fields $(\bar{\phi}_0, \bar{\phi}_1, \bar{\phi}_2, \bar{\phi}_3)$ which have the Fourier series expansions as~\cite{Lee:2003eg}~\footnote[1]{Instead of expanding with the unnormalized modes, one can also make a Fourier expansion of the normalized modes $\phi_\alpha (\alpha = 0,1,2,3)$ defined by Eq.~\eqref{eq:definenorphi}~\cite{Suenaga:2014sga}.}
\begin{equation}
\bar{\phi}_0 = \sum_{a,b,c} \bar{\beta}_{abc} \cos(a\pi x/L)
\cos(b\pi y/L) \cos(c\pi z /L),
\label{sigma}
\end{equation}
and
\begin{eqnarray}
\bar{\phi}_1 &=& \sum_{h,k,l} \bar{\alpha}_{hkl} \sin(h\pi x/L)
\cos(k\pi y/L) \cos(l\pi z/L),
\label{pi1} \\
\bar{\phi}_2 &=& \sum_{h,k,l} \bar{\alpha}_{hkl} \cos(l\pi x/L)
\sin(h\pi y/L) \cos(k\pi z/L),
\label{pi2} \\
\bar{\phi}_3 &=& \sum_{h,k,l} \bar{\alpha}_{hkl} \cos(k\pi x/L)
\cos(l\pi y/L) \sin(h\pi z/L).
\label{pi3}
\end{eqnarray}
The fields $\bar{\phi}_\alpha$ relate to the normalized field $\phi_{\alpha}$ through
\begin{eqnarray}
\phi_\alpha & = & \frac{\bar{\phi}_\alpha}{\sqrt{\sum_{\beta=0}^3 \left(\bar{\phi}_\beta\right)^2}}, \quad (\alpha,\beta = 0,1,2,3).\label{eq:norm}
\end{eqnarray}
By taking the Fourier coefficients $\bar{\beta}$ and $\bar{\alpha}$ as free parameters and varying them, the minimum energy per particle at a specific crystal size renders the quasi-nucleon density-dependent. Note that for a specific crystal structure, the Fourier coefficients $\bar{\beta}$ and $\bar{\alpha}$ are not independent, but are related. Note also that the normalization (\ref{eq:norm}) does not spoil any symmetries that the unnormalized fields possess, while the expansion coefficients $\bar{\alpha}_{hkl}$ and $\bar{\beta}_{abc}$ lose their meaning as Fourier coefficients in the normalized fields~\cite{Kugler:1988mu,Kugler:1989uc}.

Among a variety of properties revealed in the crystal approach to dense matter, the most important of all is the existence of half-skyrmion configurations at some higher density. Being topological, its presence is a robust prediction involving maximal symmetry~\cite{goldhaber-manton}.  Its does not depend on what degrees of freedom other than the pion are involved. The detailed discussions and extensive references are found in \cite{BReditor}, so we invite the interested readers to consult that volume. What is significant is that it involves a topology change from skyrmions to half-skyrmions, which is responsible for a dramatic change in the properties of the dense matter at a density $n_{1/2}\sim 2n_0$, a feature which is not visible in other approaches in the literature. The topological change is essential in the given effective field theory (EFT) for the equation of state for compressed baryonic matter relevant for massive compact stars~\cite{Harada:2016tkf}.

\begin{figure}[H]\centering
\includegraphics[scale=0.2]{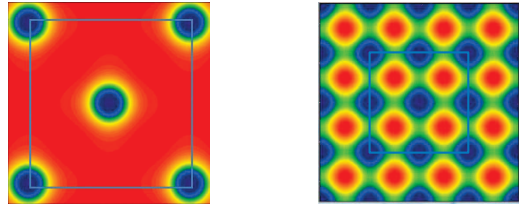}
\caption{The distribution of the baryon number density in the skyrmion (left panel) phase and half-skyrmion  phase (right panel). }
\label{fig:SkyrPhaseTran}
\end{figure}

In Fig.~\ref{fig:SkyrPhaseTran} is shown the baryon number density distribution of the skyrmion (left panel) and half-skyrmion states of matter (right panel). In the half-skyrmion phase, the vertices where the baryon number density accumulates form the CC crystal. The space average $\phi_0$ defined as
\begin{eqnarray}
\langle \phi_0 \rangle & = & \frac{1}{(2L)^3}\int^{2L}_0 d^3 x \,\phi_0 , \label{eq:defvevphi0}
\end{eqnarray}
which is nonzero in the skyrmion matter, is found to vanish in the half-skyrmion state. This implies that the quark condensate $\langle\bar{q}q\rangle$ vanishes when the space is averaged. But this does not imply a phase transition since chiral symmetry is not restored with nonvanishing pion decay constant, that is, the pion is still present. In fact there is no bona-fide order parameter for this change of state, symptomatic of topology change. Although it is not a paradigmatic phase change, topology change can be considered as a phase transition, so in what follows we will use ``half-skyrmion phase."

When the skyrmion-half-skyrmion phase transition takes place, in addition to the original lattice vertices, the baryon number density accumulates in the middle of the lines connecting the original FCC lattice vertices. If one integrates the baryon number density in the region where baryon number density accumulates, each new crystal lattice has baryon number one-half. At what density the transition can take place in dense baryonic matter depends on, among others, the degrees of freedom that intervene. Since in practice a large number of mesons can be involved, e.g., the infinite tower of holographic QCD, and one is limited to drastic approximations, it is difficult to pinpoint the changeover density $n_{1/2}$. Given that there is no indication {for  half-skymrion phase} at nuclear matter density, it cannot be lower than the normal density $n_0$. If it were much higher, say, in the regime where the percolation transition can take place, then the model could not be used. Thus the optimal density for $n_{1/2}$ should be somewhere above $n_0$. The present estimate based on detailed analyses using a semi-realistic model that contains $\pi$, $\rho$, $\omega$ and $\sigma$ indicates that $n_{1/2} \sim 2n_0$.  It is this density regime where hadron-quark continuity seems to take place as we will mention in Sec.~\ref{sec:power}. We will keep this as a typical value in applications to nuclear matter and denser matter.

It should be mentioned before we enter into detailed discussions that the skyrmion crystal picture would make sense if $N_c$ were very large. Indeed if $N_c\to \infty$, then the baryonic matter would be a crystal. In reality $N_c=3$ is not so ``large" and normal nuclear matter is a liquid and not a crystal. One might hope to arrive at a structure compatible with nuclear matter by doing proper quantum-mechanical calculations taking into account $1/N_c$ corrections starting from the crystal structure but nobody succeeded to do such calculations so far. Even within the large $N_c$ scheme, indication from holographic QCD calculations is that at high density, a variety of different forms, such as dyonic salt (resembling half-skyrmions), layers of ``popcorns," salty or sugared etc, can be present~\cite{Rho:2009ym,Gorsky:2013dda,Bolognesi:2013nja,Kaplunovsky:2012gb,Elliot-Ripley:2016uwb,Elliot-Ripley:2016ctk}. All these exotic configurations could be artifacts of the large $N_c$ models but at present none of them can be ruled out.

Given this wilderness of scenarios, what  we will adopt is the topology change of skyrmion-half-skyrmion phase change and exploit it in both the crystal approach and continuum approach that takes into account the topology change in the parameters of the effective Lagrangian.  We will eschew relying on the numerical results coming from the crystal lattice.

\section{Dense Matter in the Skyrme Model}

\label{sec:power}

We first discuss certain generic properties of the skyrmion approach in terms of the simplest model, the Skyrme model (with pions only). The Skyrme model may be considered as a simplified chiral Lagrangian with all heavy degrees of freedom than pion integrated out and all higher derivative terms dropped.   The resulting Lagrangian contains the current algebra term and the quartic term, i.e., the Skyrme quartic term. In fact the quartic term is essentially the leading term in the next-to-leading order $O(p^4)$ chiral Lagrangian that captures $\pi$-$\pi$ scattering~\cite{diakonov-petrov}. For simplicity, we take the parameters given by the free-space value rather than the effective values fitting nucleon properties or nuclear properties used in the literature. Here the purpose is to obtain a qualitative idea of what's going on rather than fitting data. To have a better view of what we are driving at, it is useful to recall the currently popular approach anchored on phenomenological Lagrangian to finite nuclei, nuclear matter and dense matter relevant to compact stars, namely, the energy-density functional approach (EDFA for short). Related closely to what we are doing is the relativistic mean-field (RMF) theory, initiated by Walecka~\cite{Walecka:1974qa}. This theory works fairly well up to the density appropriate for nuclear matter. This can be understood in terms of Landau Fermi-liquid theory, to which the RMF at large density is related.

A typical phenomenological Lagrangian used is of the form
%\begin{widetext}
\begin{eqnarray}
{\cal L}& = & \bar{\psi}\left( i \gamma_\mu \partial^\mu - m_N \right)\psi \nonumber\\
& & {} + \frac{1}{2}\left(\partial_\mu s \partial^\mu s - m_s^2 s^2\right) \nonumber\\
& & {} + \frac{1}{2}\left(\partial_\mu \bm{a}_0 \cdot \partial^\mu\bm{a}_0 - m_{a_0}^2 \bm{a}_0^2 \right) \nonumber\\
& & {} - \frac{1}{4}\omega_{\mu\nu}\omega^{\mu\nu} - \frac{1}{2}m_\omega^2 \omega_\mu \omega^\mu \nonumber\\
& & {} - \frac{1}{4}\bm{\rho}_{\mu\nu}\cdot\bm{\rho}^{\mu\nu} - \frac{1}{2}m_\rho^2 \bm{\rho}_\mu \cdot \bm{\rho}^\mu \nonumber\\
& & {} + g_s \bar{\psi}s \psi + g_{a_0}\bar{\psi}\bm{\tau}\cdot \bm{a}_{0}\psi - g_\omega \bar{\psi}\gamma_\mu \omega^\mu \psi \nonumber\\
& & {} - g_\rho \bar{\psi}\gamma_\mu \bm{\tau}\cdot \bm{\rho}^\mu \psi - e \bar{\psi}\gamma_\mu A^\mu \psi - \frac{1}{4}F_{\mu\nu}F^{\mu\nu},
\end{eqnarray}
%\end{widetext}
 where $\psi = (p, n)^T$ are the nucleon doublet, $\rho$ is the iso-vector vector meson, $\omega$ is the iso-scalar vector meson, $s$ is the  iso-scalar scalar meson, and $\bm{a}_0$ is the iso-vector scalar meson. Included is the photon field $A_\mu$.  This Lagrangian, what one might call ``linear meson model," can be related to chiral-scale Lagrangian discussed below if the iso-vector scalar mesons are ignored. In the mean field, this Lagrangian  yields a broadly successful description of nuclear matter~\cite{Serot:1997xg}.  There are certain properties, such as compression modulus, that are at odds with experiments in this linear model.  When higher-dimension meson fields are incorporated, consistently with the given symmetries, this approach can post-dict $\sim 2000$ nuclear spectra with deviation $\leq 0.5$ MeV at the expense of $\sim 50$ parameters.

\begin{figure}[H]\centering
\includegraphics[scale=0.27]{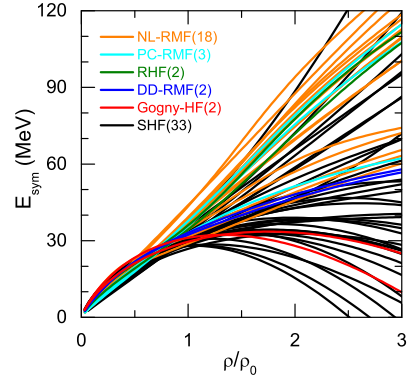}
\caption{Symmetry energy as a function of density. This illustrates that one can obtain anything for the symmetry energy at density $n > n_0$ by adjusting parameters of the mean-field or equivalently energy-density functional without affecting the normal-matter properties. This figure is copied from Ref.~\cite{Chen:2015gba}. }
\label{fig:SymmE}
\end{figure}

It will be discussed below that within the spirit of mean field approximations, one can formulate an EFT  with a Lagrangian that combines both chiral symmetry and scale symmetry with the parameters of the Lagrangian matched to the correlators of QCD.  The power of this approach will be to make predictions at high density that are difficult to access without ambiguity. The phenomenological approach, successful up to nuclear matter density, can go wild, with the numerous parameters at the disposal that are unconstrained at high density with experimental guidance. A typical example is given by the nuclear symmetry energy that strongly controls the equation of state for compact-star matter. One can see this in  Fig.~\ref{fig:SymmE}.

The power of the skyrmion crystal approach is that  this wildness can be cleared up, thanks to the topology change that unifies baryon and nuclear structure and properties of baryons in medium, obtaining a unique prediction.  This feat is the objective of what's described below.

\subsection{Nuclear matter from the skyrmion crystal model}

In this and following subsections we discuss generic features of the skyrmion model that follow from topology involved with pion field. Other massive degrees of freedom that will figure in the dynamics will influence the detailed structure without affecting the characteristics of the model.  Here we first look at nuclear matter. In doing this we employ the Skyrme model  following Ref.~\cite{Lee:2003aq}. The Lagrangian is
\begin{equation}
{\cal L}= -\frac{f_\pi^2}{4} \mbox{Tr} \left(U^\dagger \partial_\mu U
U^\dagger \partial^\mu U\right) + \frac{1}{32e^2} \mbox{Tr}
\left [U^\dagger \partial_\mu U, U^\dagger \partial_\nu U\right]^2.
\label{eq:Lsk}
\end{equation}
Since we are aiming at a qualitative rather than quantitative description of nature, we will not take effective  parameters fit to the structure of baryon and baryonic matter as was done in the literature, but will take what's determined in the vacuum. Thus we take $f_\pi$ to be the free-space pion decay constant $\simeq 93$ MeV. For the Skyrme parameter $e$, we take $e = g = 5.93$ with $g$ being the value of the gauge coupling constant of the HLS, the chiral effective theory of vector mesons~\cite{Bando:1984ej,Harada:2003jx}, which may be obtained by assuming the $\rho$ meson mass is ``heavy" \footnote{ This assumption must break down at large density approaching the vector manifestation fixed point.}. The robust, though qualitative, properties extracted therefrom will then be the basis for doing more realistic calculations including the other relevant degrees of freedom.

\begin{figure}[H]\centering
\includegraphics[scale=0.25]{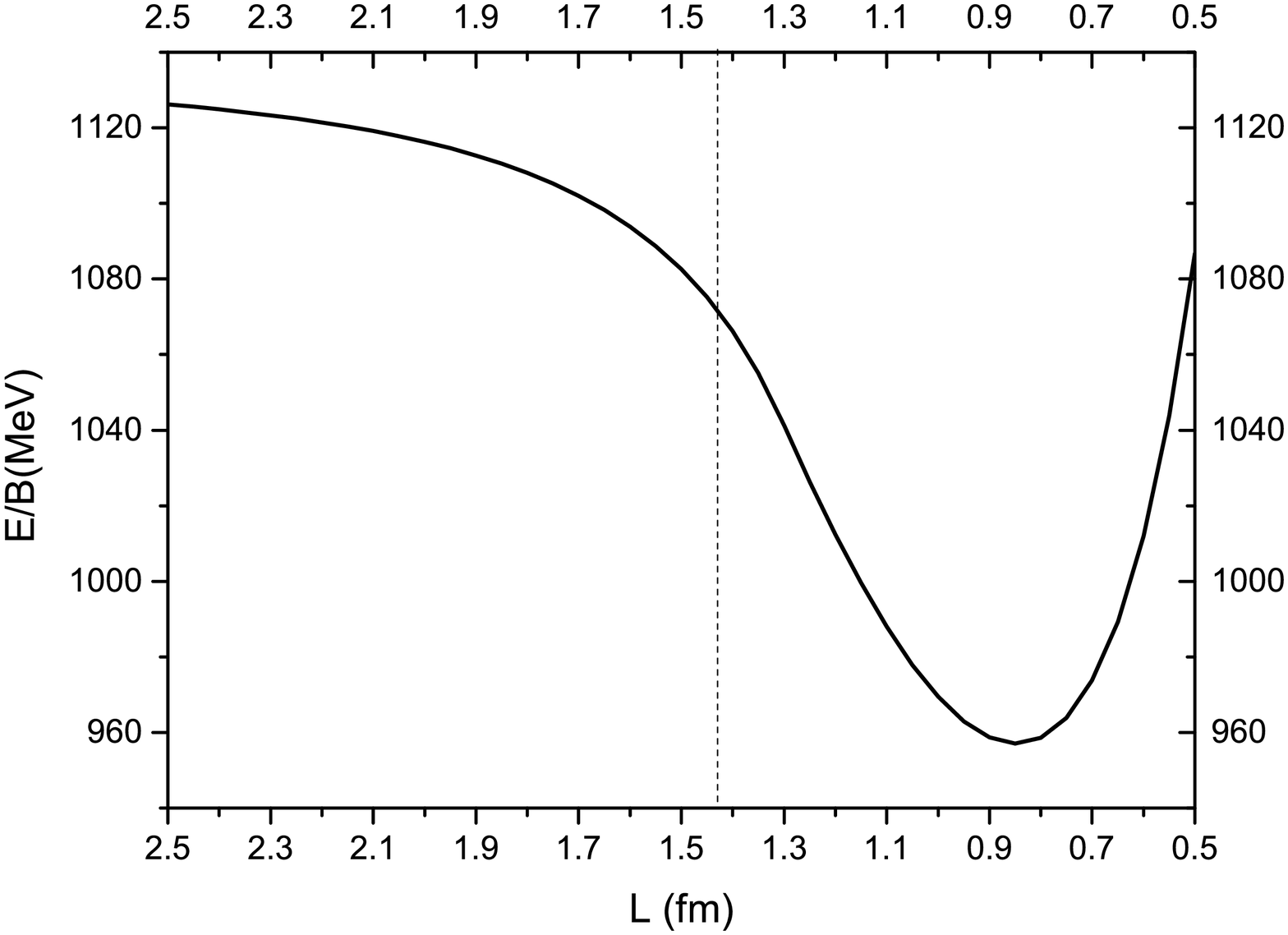}
\includegraphics[scale=0.23]{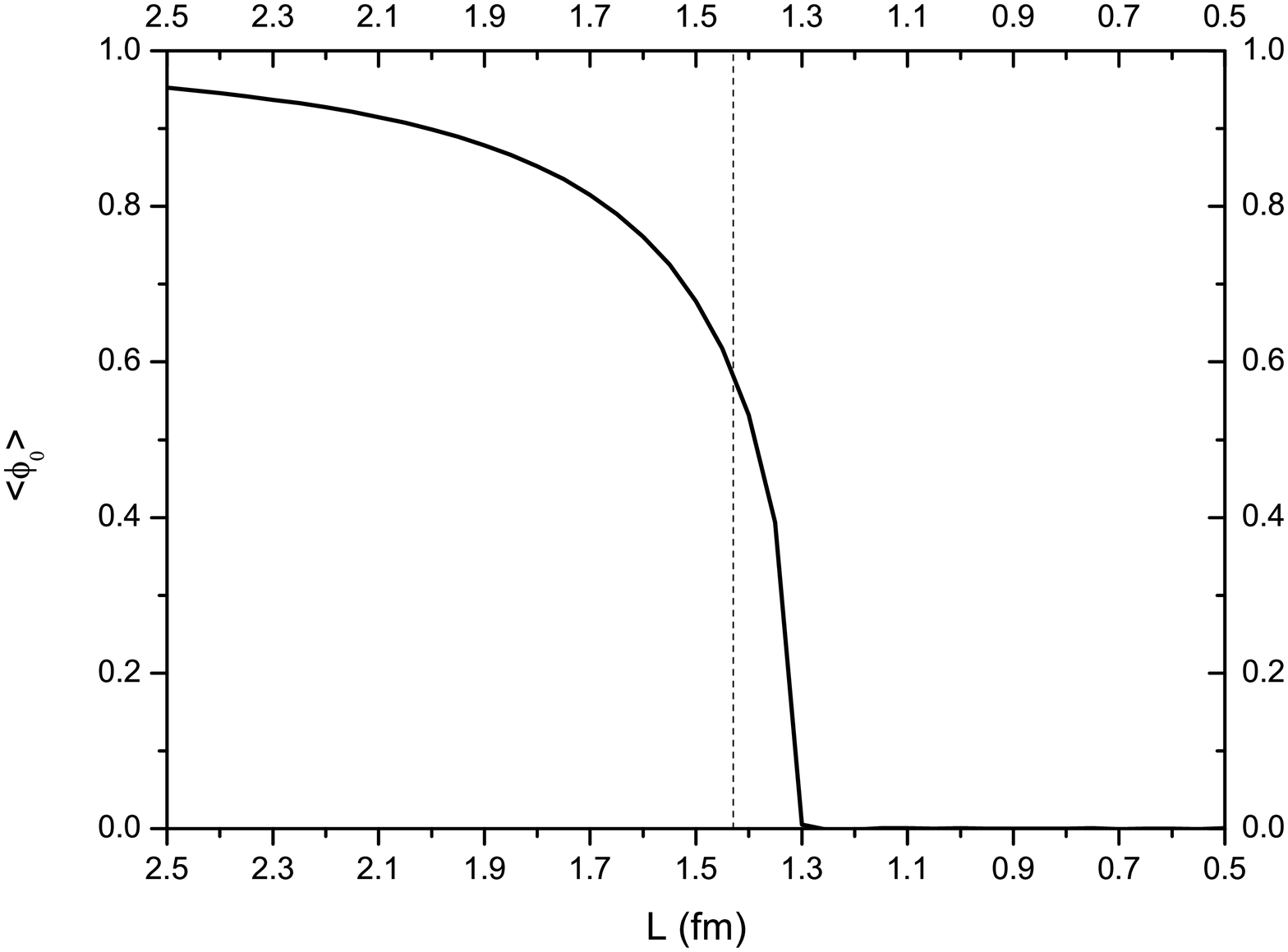}
\caption{Per-skyrmion energy $E/B$ (upper panel) and $\langle\phi_0\rangle$ (lower panel) as a function of crystal size $L$.}
\label{fig:fccfuncL}
\end{figure}

By putting the Skyrme solitons of \eqref{eq:Lsk} onto the FCC crystal so as to simulate  nuclear matter, one obtains the result of  Fig.~\ref{fig:fccfuncL} for the skyrmion-matter per particle (or ``per-skyrmion") energy  $E/B$ and $\langle\phi_0\rangle$ -- which is related to the quark condensate in QCD variables through decomposition ~\eqref{eq:definenorphi} -- as a function of crystal size $L$. The vertical line indicates the crystal size that corresponds to  the normal nuclear density.  One notes that the system has a minimum at a density considerably higher than what is normal density. This is not surprising. It just indicates that the Skyrme Lagrangian written to the lowest order chiral counting is missing some important components in nuclear interactions which give a net repulsive force. In fact what corresponds in standard chiral perturbative approaches to multi-body (i.e., three-body) repulsive force -- of the range of an $\omega$-exchange force that stabilizes nuclear matter --  is absent. Also the resulting large binding energy, $\sim 200$ MeV, is not surprising given that the typical energy scale involved in QCD is the $\Lambda_{QCD}$, which is a few times the pion decay constant $f_\pi\approx 93$ MeV. All these features are understandable from the large $N_c$ counting, on which the skyrmion model is relying. These defects can be remedied in refined, though involved, treatments. In fact when vector mesons and scalar mesons are included, the situation improves substantially~\cite{Ma:2012kb,Ma:2012zm,Ma:2013ooa,Ma:2013ela} and with an infinite tower one obtains even agreement with experiments~\cite{Sutcliffe:2010et,Sutcliffe:2011ig,Ma:2012kb}. We will return to this matter below.

On the contrary, what is highly non-trivial and of great importance that can be exploited in our more realistic approach is the observation that the quantity $\langle\phi_0\rangle$ representing the quark condensate vanishes for density above normal nuclear density. At what precise density that takes place cannot be pinned down. But it is a robust observation. One observes in Fig.~\ref{fig:fccfuncL} that when the system is squeezed from $L=2.5$\,fm to around $L=1.3$\,fm, the skyrmion system undergoes a phase transition from the FCC skyrmion configuration to the CC half-skyrmion configuration. The system has a minimum energy at $L\sim0.85$\,fm
with the energy per baryon $E/B\simeq 957$\,MeV. Of significance is that the value of  $\phi_0$  averaged over space rapidly drops as the system shrinks. It reaches zero at $L\sim 1.3~$fm where the system goes to a half-skyrmion phase mentioned above. This ``phase transition" can be interpreted, once the pion fluctuations are incorporated, as a signal for {\it global} chiral symmetry restoration whereas {\it locally} the system is still in the chiral symmetry broken phase~\footnote{The density at which the skyrmion matter transits to the half-skyrmion phase obtained here is different from that obtained in Ref.~\cite{Lee:2003aq} due to different Skyrme parameters. If we took the empirically fit value $e = 4.75$, the skyrmion to half-skyrmion phase transition would appear above the normal nuclear density. One should not take the numerical values we are obtaining too seriously.}.  In this description there is no hint of QCD variables, i.e.,  quarks and gluons. However the skyrmion-half-skyrmion transition can be reproduced in terms of a chiral-quark model where the transition is captured in a delocalization of the baryon number density leading to $b=1/2$ structure~\cite{MantovaniSarti:2014nqa}. There is another transition in this picture corresponding to quark deconfinement at higher density, which is however not in the  half-skyrmion phase.

\subsection{In-medium pion decay constant}

One of the most crucial quantities in dense medium is the effective in-medium pion decay constant $f_\pi^{\ast}$. It controls medium modified hadron properties~\cite{Lee:2003aq,Lee:2003eg}. Here, as a concrete example, we calculate the medium-modified pion decay constant $f_\pi^{\ast}$, again taking the Skyrme model. The latter contains the essential feature that is shared by more realistic models given below.

The procedure is quite simple. We take the skyrmion crystal solution as the background classical field and interpret the fluctuating fields on top of it as the corresponding mesons in
dense baryonic matter. Following Ref.~\cite{Ma:2013ela}, we write the skyrmion crystal solution as $U_{(0)}$ and introduce the fluctuating fields as
\begin{eqnarray}
U & = & u_{(0)}\check{U}u_{(0)}, \label{eq:fieldfluct}
\end{eqnarray}
where $\check{U} = \exp (2i \tau_a \check{\pi}_a/f_\pi)$ stands for the corresponding fluctuating field and $u_{(0)}^2 = U_{(0)}$. Since for each crystal size, there is a solution for $U_{(0)}$, $U_{(0)}$ is dependent on the crystal size, or equivalently, the baryon density. The decomposition in Eq.~\eqref{eq:fieldfluct} guarantees that to each order of the fluctuation the chiral invariance of the model is preserved. By substituting the fields in Eq.~\eqref{eq:fieldfluct} into the Skyrme model Lagrangian one obtains the medium-modified Lagrangian for the pion. Consequently the parameters in the pion Lagrangian become density-dependent through their dependence on $U_{(0)}$.

\begin{figure}[H]\centering
\includegraphics[scale=0.6]{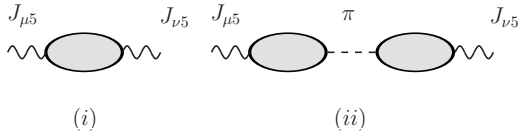}
\caption{Two types of contributions to the correlator of
Eq.~\eqref{eq:defaacorr}: (i) the contact diagram and (ii) the pion exchange diagram. Shaded blobs stand
for interaction vertices in the skyrmion matter.  }
\label{fig:Correlator}
\end{figure}

To define the pion decay constant in the skyrmion matter, we write the axial-vector current correlator
\be
iG_{\mu\nu}^{ab}{(p)} = i\int d^4x e^{ip\cdot x}\la 0|T J_{5\mu}^a{(x)} J_{5\nu}^b{(0)} |0\ra\label{eq:defaacorr}
\ee
 which is decomposed in medium as~\cite{Kapusta-book}
\begin{eqnarray}
G_{\mu\nu}^{ab}(p) &=& \delta^{ab} \left[ P_{T\mu\nu} G_{T}(p)
+ P_{L\mu\nu} G_{L}(p) \right],
\end{eqnarray}
where the polarization tensors $P_{L,T}$ are defined as
\begin{eqnarray}
P_{T\mu\nu}^{} &=&
g_{\mu i}^{} \left( \delta_{ij} - \frac{p_i^{} p_j^{}}{|\bm{p}|^2} \right) g_{j\nu} ,
\nonumber\\
P_{L\mu\nu}^{} &=& - \left( g_{\mu\nu} - \frac{p_\mu^{} p_\nu^{}}{p^2}
\right) - P_{T\mu\nu}^{}.
\end{eqnarray}
Then the medium-modified (time component of) pion decay constant is given by the longitudinal
component in the low-energy limit
\begin{eqnarray}
f_\pi^{\ast 2} & \equiv & {} - \lim_{p_0 \to 0} G_{L}(p_0,\bm{p}=0) .
%\label{eq:mediumfpi}
\end{eqnarray}

Now for the calculation, we shall ignore the contributions from the loop diagrams of the fluctuation fields. Therefore, there are only two types of diagrams that contribute as illustrated in Fig.~\ref{fig:Correlator}: (i) the contact diagram and (ii) the pion exchange diagram. They are given by
\begin{eqnarray}
&\mbox{(i)} : & i f_\pi^2 g_{\mu\nu}^{} \delta^{ab}
\left(1-\frac{2}{3}\left\langle\bm{\phi}_\pi^2\right\rangle \right) ,
\nonumber\\
&\mbox{(ii)} : &
- i f_\pi^2 \frac{p_\mu^{} p_\nu^{}}{p^2} \delta^{ab}\left(1-\frac{2}{3}\left\langle\bm{\phi}_\pi^2\right\rangle \right) .
\end{eqnarray}
Summing the two contributions, one obtains
\begin{eqnarray}
f_\pi^{\ast 2} & = & {} f_\pi^2 \left [1 - \frac{2}{3}
\left( 1-   \left\langle {\phi_0^2}\right\rangle \right)\right] ,
\label{eq:mediumfpi}
\end{eqnarray}
where the intrinsic density dependence is brought in by the minimal energy solution $\phi_0^2$.  To arrive at ~\eqref{eq:mediumfpi}, we used the relation
$\phi_0^2  + {\bm \phi}_\pi^2 = 1$. Equation \eqref{eq:mediumfpi} shows the direct relation between the medium-modified pion decay constant $f_\pi^{\ast}$ and the parameter $\langle\phi_0\rangle$ which signals the phase transition involving topology.

\begin{figure}[H]\centering
\includegraphics[scale=0.27]{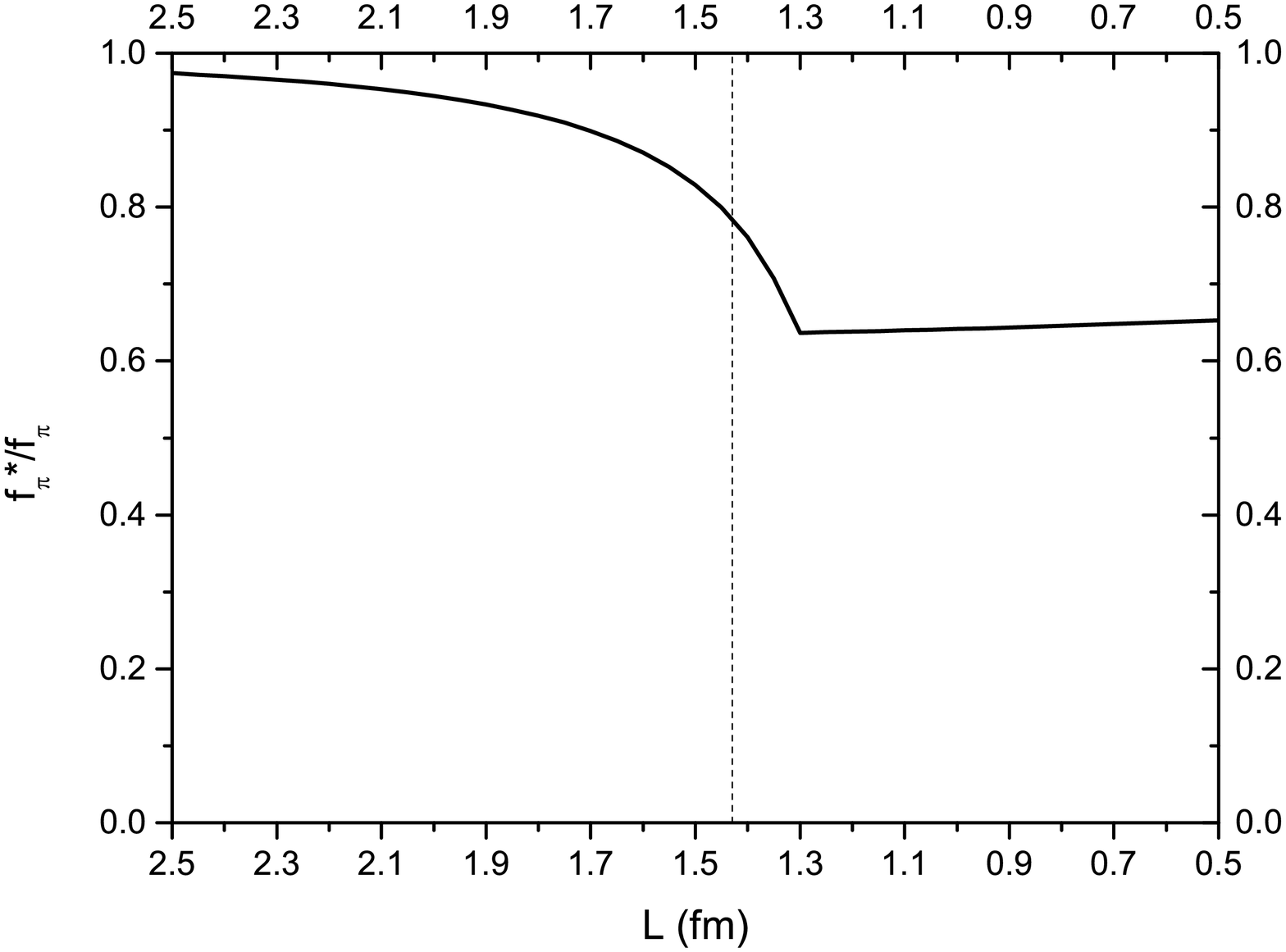}
\caption{$f^*_\pi/f_\pi$ as a
function of the crystal size $L$ .}
\label{fig:inmediumpion}
\end{figure}

In Fig.~\ref{fig:inmediumpion} is plotted the crystal-size dependence of $f^*_\pi/f_\pi$. We see that $f^*_\pi$ decreases to $\sim 0.65 f_\pi$ at the density $n_{1/2}$ at which skyrmions fractionize to half-skyrmions  and then stays constant for $n > n_{1/2}$. In terms of the quark condensate $\la\bar{q}q\ra$, $\langle \phi_{0}\rangle$  corresponds to the space average of the latter. It should be noted that $\langle \phi_{0}\rangle \to 0$  does not signal the restoration of chiral symmetry since the pions are still around with the non-zero pion decay constant. This means that the quark condensate is locally non-zero in the half-skyrmion phase, implying that there is chiral-density wave. This was  verified in ~\cite{Harada:2015lma}. This feature will remain unmodified  when scale symmetry is implemented. This results from the locking of scale symmetry and chiral symmetry, consistent with the scale-chiral effective field theory~\cite{CT} -- discussed below --  with the consequence that $f_\pi^\ast\approx f_\sigma^\ast$.

\subsection{Kaons in dense matter}

Another highly non-trivial prediction of skyrmion crystal matter is the property of kaons in medium.  Kaon property in nuclear and denser matter has been an important issue for the equation of state of compact stars~\cite{Brown:2007ara} since kaon condensation was predicted on the basis of chiral symmetry in 1985~\cite{Kaplan:1986yq}. The issue is still far from being settled and is becoming an urgent matter, in particular with the discovery of massive neutron stars.  We will discuss here what one can say about the properties of kaons in dense nuclear matter from the skyrmion crystal point of view. Again we will use the Skyrme model extended to three flavors including strangeness, thus involving octet NG bosons, but minimally implemented with  dilaton dynamics. Although highly simplified, the model we use exposes a non-trivial effect of the dilaton field that is absent in the  usual chiral Lagrangian approaches.

Following \cite{Park:2009mg}, we take the Lagrangian
\begin{eqnarray}
{\cal L}_{\rm K} & = & \left(\frac{\chi}{f_\chi}\right)^2D_\mu K^\dagger D^\mu K - \left(\frac{\chi}{f_\chi}\right)^3m_K^2 K^\dagger K \nonumber\\
& &{} - \frac{2iN_c}{5f_\pi^2} {B}^\mu \left(D_\mu K^\dagger K - K^\dagger D_\mu K\right),
\label{eq:KaonSkyrmion}
\end{eqnarray}
where $K$ is the kaon doublet with $K = (K^+ , K^0)$, the last term is the topological Wess-Zumino term present for three-flavor systems,  the covariant derivative is defined as
\begin{eqnarray}
D^\mu K  = (\partial^\mu + \Gamma^\mu)K ,
\end{eqnarray}
where
\begin{eqnarray}
\Gamma^\mu & = & \frac{1}{2}\left[u^\dagger \partial^\mu u + u \partial^\mu u^\dagger\right],
\end{eqnarray}
with $u \equiv\sqrt{U}$ and the baryon number current as
\begin{eqnarray}
B^\mu & = & \frac{1}{24\pi^2}\epsilon^{\mu\nu\alpha\beta}{\rm tr}\left[u^\dagger \partial_\nu u\cdot u^\dagger \partial_\alpha u\cdot u^\dagger \partial_\beta u \right].
\end{eqnarray}
In Eq.~\ref{eq:KaonSkyrmion}, the dilaton field $\sigma$ is introduced via what is called ``conformal compensator" $\chi$ which has mass dimension-1 and scale dimension-1,
\be
\chi=f_\sigma e^{\sigma/f_\sigma}.
\ee
This is because the dilaton field plays a specially important role.  How this field enters in chiral symmetry Lagrangians to join in  chiral-scale symmetry will be described later. Here it suffices to say that the Lagrangian accounts for scale symmetry transformation to the leading order of scale-chiral counting rule~\cite{CT,Li:2016uzn}.

The kaon field is affected by the background of nuclear matter through the skyrmion profile $u$.  Taking the mean-field approximation in  \eqref{eq:KaonSkyrmion}, one obtains the kaon dispersion relation in the skymion matter as
\begin{eqnarray}
\alpha \left(\omega_K^2 - \vec{p}_K^2\right) + 2\beta \omega_K + \gamma & = & 0 ,
\end{eqnarray}
where
\begin{eqnarray}
\alpha & = & \frac{1}{4}\left\langle\left(\frac{\chi}{f_\chi}\right)^2 \left(u + u^\dagger + 2\right) \right\rangle, \nonumber\\
\beta & = & \frac{N_c}{16f_\pi^2}\left\langle B_0 \left(u + u^\dagger + 2\right) \right\rangle, \nonumber\\
\gamma & = & \frac{1}{4}\left\langle \left(\frac{\chi}{f_\chi}\right)^3 \left(u + u^\dagger + 2\right) \right\rangle m_K^2.
\end{eqnarray}
Focusing on S-wave kaons with $\vec{p}_K \to 0$, the effective in-medium kaon energy is obtained by solving the equation
\begin{eqnarray}
m_K^\ast \equiv \lim_{\vec{p}_K \to 0} \omega_K = \frac{-\beta + \sqrt{\beta^2 + \alpha \gamma}}{\alpha} .
\end{eqnarray}

\begin{figure}[H]\centering
\includegraphics[scale=0.25]{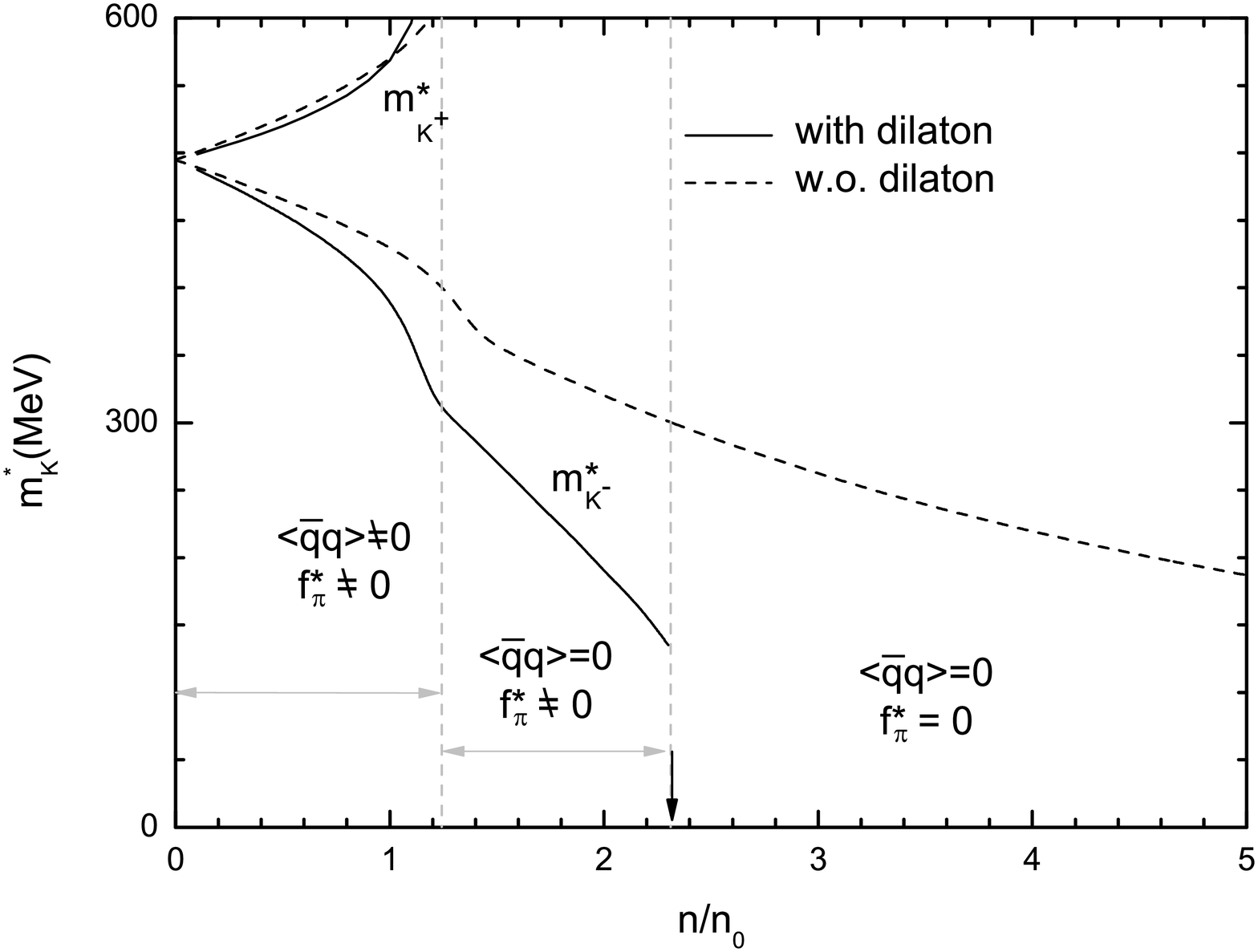}
\caption{Skyrmion matter modified Kaon mass $m_K^\ast$ as a function of density $n/n_0$ for dilaton mass $m_\chi = 600~$MeV. See Ref.~\cite{Park:2009mg} for an explanation of the figure.}
\label{fig:inmediumKaon}
\end{figure}

The medium-modified kaon mass $m_K^\ast$  (energy for the S-wave) is shown in Fig.~\ref{fig:inmediumKaon} which is borrowed from Ref.~\cite{Park:2009mg}.  The effect of $\la\phi_{0}\ra\to 0$ toward $n_{1/2}$ is visible for both with and without the dilaton. However the dropping of  $m_K^\ast$ is accentuated by the presence of the dilaton. Furthermore the mass drops at a faster rate in the half-skyrmion phase. By itself, this would have a big effect on the process of kaon condensation in compact star matter. It should be noted that certain baryonic short-range correlations, missing in mean-field-type considerations (appropriate for Fermi-liquid fixed-point treatments), should be included for realistic calculations. This could apply both to kaon condensation and to hyperon interactions in compact star matter. This would require an approach drastically different from the naive application of chiral perturbation theory. We will see later such an effect could have a strong impact on the formation of dense kaonic matter at the hadronization phase in the cosmological evolution of the Universe~\cite{yamazaki}.

\subsection{The nuclear symmetry energy}

Perhaps the most important impact of the skyrmion-half-skyrmion (topological) phase change in the physics of dense matter is on the nuclear symmetry energy that has defied theorists in compact-star physics since many years as discussed above (See Fig.~\ref{fig:SymmE}). It turns out  that it is tightly connected to the vector manifestation~\cite{Harada:2000kb} for the $\rho$ meson in the skyrmion approach and it is not at all clear how that translates into standard effective field approach based on chiral symmetry. Quite remarkably it offers a compelling evidence that the topology change, apparently absent in many-body interactions, is robust and can be exploited to construct a realistic model for dense matter that cannot be accessed by other approaches. It turns out that the change from skyrmions to half-skyrmions plays a key role.

The energy per particle $E(n,\delta)$ of a dense system with density $n$ and  asymmetry $\delta$ can be written as
\be
E(n,\delta)=E_0(n,x=0) +\delta^2 E_{sym}(n) +O(\delta^4)+\cdots\, ,
\ee
where $\delta=(N-P)/A$ with $N(P)$ being the neutron (proton) number and $A=N+P$. This equation defines the symmetry energy $E_{sym}$. It turns out that the above expansion is a good one.  The corrections entering at $O(\delta^2)$ are estimated to be quite small even for $\delta$ near 1. We will ignore them here.

In the skyrmion crystal approach, the symmetry energy $E_{sym}$ can be calculated for the neutron matter ($\delta=1$) for large $N$ by collective-rotating the whole system~\cite{Li:2016uzn,Lee:2010sw}. After the collective rotation, the energy of the $A$-nucleus is
\begin{eqnarray}
E_{\rm A} & = & A E_{\rm Sol} + \frac{1}{2 A {\mathcal I}}I^{\rm Tot}\left(I^{\rm Tot} + 1 \right) ,
\end{eqnarray}
where $E_{\rm Sol}$ is the per-soliton energy in the nucleus and  $A{\cal I}$ is the iso-spin moment of inertia of the $A=N$  system. $I^{\rm Tot}$ is the total isospin of the whole system which could be the same as the third component of the isospin $I_3$ for pure neutron matter, i.e., $I^{\rm Tot} = A/2$. For $\delta \equiv (N-P)/(N+P) \lesssim 1$ we have
\begin{eqnarray}
I^{\rm Tot} & = & \frac{1}{2}\left(N - P\right) = \frac{1}{2}\left(N + P\right)\frac{\left(N - P\right)}{\left(N + P\right)} = \frac{1}{2}A \delta.
\end{eqnarray}
The energy per nucleon can then be expressed as
\begin{eqnarray}
E_{\rm Nucleon} & = & \frac{E_{\rm Nuclei}}{A} \simeq E_{\rm Sol} + \frac{1}{8 {\cal I}}\delta^2 ,
\end{eqnarray}
for a large mass number $A$. Consequently, the symmetry energy takes the simple form
\begin{eqnarray}
E_{\rm Sym} & = & \frac{1}{8 {\cal I}}.
\end{eqnarray}

\begin{figure}[H]\centering
\includegraphics[scale=0.27,angle=270]{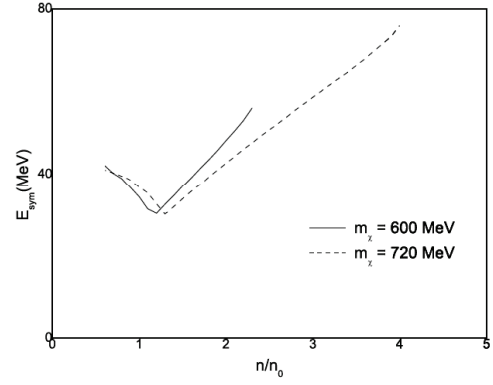}
\caption{Symmetry energy as a function of density calculated in Ref.~\cite{Lee:2010sw}. The cusp is located at $n_{1/2}$. The low
density part that cannot be located precisely is not shown as
the collective quantization method used is not applicable in
that region.}
\label{fig:SymmESkyr}
\end{figure}

In Ref.~\cite{Lee:2010sw}, using the Skyrme model implemented with scale symmetry, $E_{\rm Sym}$ was calculated on crystal lattice in the way described above. The result  is shown in Fig.~\ref{fig:SymmESkyr} copied from Ref.~\cite{Lee:2010sw}. The symmetry energy first decreases with density up to the density $n_{1/2}$ at which skyrmions fractionize to half-skyrmions and then increases as density increases. This is the appearance of the cusp in  Fig.~\ref{fig:SymmESkyr}.

Of high significance is that, strange or anomalous though it may appear, the presence of the cusp, arising from the skyrmion-half-skyrmion topology change involving the pion field,  is robust and does not depend on what other degrees of freedom are incorporated in the dynamics. It turns out, as pointed out below, that the cusp plays an essential role in the EoS of massive compact stars~ \cite{Dong:2012ch,Paeng:2015noa}.

A question that is immediately raised is whether this cusp is not just an artifact of the crystal structure. We address this question and show that there is neither abnormality nor mystery in the cusp. How this cusp arises can be explained in terms of standard nuclear many-body interactions involving the nuclear tensor forces coming from the exchanges of $\pi$ and $\rho$ whose properties are influenced by the topology change. Quite independently of whether it is valid or not, it is a new phenomenon hitherto undiscovered in the field.

\begin{figure}[H]\centering
\includegraphics[scale=0.27]{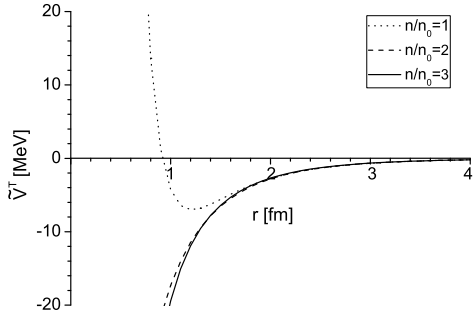}
\caption{Tensor force as a function of density. For definiteness $n_{1/2}=2n_0$ is taken.}
\label{fig:tensor}
\end{figure}

The treatment is much more straightforward with HLS Lagrangian with nucleon fields put in explicitly rather than generated as skyrmions. To address the nuclear symmetry energy, the key ingredient is the nuclear tensor force, by which the symmetry energy is dominated. Thus we can focus on the structure of nuclear tensor forces. With nucleons suitably coupled to HLS fields, the tensor force is given by the pion exchange and the $\rho$ exchange. One can ignore higher resonances for the latter. The $\pi$ and $\rho$ tensor forces have the same spin-isospin operator and radial form, differing in the overall coefficients with opposite signs and the dependence of mass. With the sign difference, the pion tensor gets reduced by the $\rho$ tensor in the intermediate and long range that the potentials act. Now since the $\rho$ mass scales with density {because of the density dependence of $f_\pi^\ast$},  whereas the pion mass remains unscaled, the $\rho$ tensor, becoming stronger at increasing density, cancels the pion tensor, almost completely, near $n_{1/2}$.  However at $n_{1/2}$, due to the topology change, the $\rho$ tensor gets strongly suppressed because the hidden gauge coupling drops rapidly due to the vector manifestation. For $n\geq  2n_0$,   the  pion tensor takes over.  This can be seen in Fig.~\ref{fig:tensor} borrowed from~\cite{Lee:2010sw}.

Although for a fair description it would require a lot more detailed discussion involving ingredients that are not touched upon in this review, such as renormalization-group treatment, it is perhaps proper to mention that the property of the tensor force in the density regime up to $n_0$ accessible experimentally is verified in the C14-dating Gamow-Teller transition~\cite{C14}. There are different ways of explaining the same phenomena but they are not alternatives but in some sense equivalent in physics.

Assuming that the tensor force dominates the symmetry energy -- which is supported also by various conventional nuclear models  -- and given that the tensor force excites states predominantly to those peaked at  $\bar{E}\sim 200$ MeV, the symmetry energy can be approximately given by the closure formula~\cite{Brown:1994pq}
\be
E_{sym}\approx \frac{\left\la \left|V^T\right|^2\right\ra}{\bar{E}}.\label{cuspform}
\ee
From Fig.~\ref{fig:tensor}, it is immediately seen that $E_{sym}$ will decrease as density approaches $n_{1/2}$ from below and then increase, reproducing the cusp at the transition density of the skyrmion crystal calculation, Fig.~\ref{fig:SymmESkyr}.

Two comments are in order regarding the cusp structure. First,   $E_{sym}$ given by (\ref{cuspform}) will, in nature, be smoothed by higher-order nuclear correlations. Indeed a refined formulation based on renormalization-group flow flattens, as will be shown below,  the cusp into a changeover from a soft to hard EoS at $n_{1/2}$ via the symmetry energy. Second, the cusp form given by (\ref{cuspform}) provides a strong support to the topology change taking place at $\sim 2 n_0$ with a qualitative effect at that density.

\section{Skyrmion Matter in Holographic QCD and Infinite Tower of Vector Mesons}

\label{sec:holography}

We have seen that  the minimal skyrmion crystal approach using the pion-only Skyrme model, with or without a  scalar degree of freedom, simulates, semi-quantitatively correctly, certain characteristic properties of nuclear matter.  But there are certain other properties difficult to access in the standard chiral effective theory approaches, such as the symmetry energy at high density, that un-mistakenly indicate massive non-NG bosons can be very important. In this section we explore the effect of the vector resonances -- including the infinite tower  inspired by the holographic model of QCD -- on dense skyrmion structure.

As mentioned above, a powerful -- and rather mysteriously efficient~\cite{komargodski} -- way of capturing  the physics of vector mesons is to resort to HLS~\cite{Bando:1984ej,Harada:2003jx}.  It is also in the HLS approach that the possibility of the $\rho$ mass approaching the zero-mass (in the chiral limit) pion at high density can be treated systematically~\cite{Harada:2003jx}.  When the energy scale goes up, the number of redundancies, therefore the number of gauge bosons, increases. The infinite number of hidden gauge vector fields that arise from the redundancies together with the pion field in 4D can be dimensionally de-constructed to 5D Yang-Mills (YM) action in curved space~\cite{Son:2003et} with the 5th (holographic) direction representing  energy scale plus the Chern-Simons term encoding chiral anomalies.

The construction of the holographic models of QCD has two approaches, what one might call "top-down" and the other "bottom-up". In the "top-down" approach, the action of the 5D Yang-Mills is reduced from the gravity sector, that is the ``bulk" sector of
gravity/gauge (holographic) duality coming from string theory. Among a variety of models given in the bulk sector, the so-called Sakai-Sugimoto (SS) model~\cite{Sakai:2004cn,Sakai:2005yt} is the one which has the symmetry properties closest to QCD. When the model is dimensionally reduced to 4D \`a la Kaluza-Klein, it has an infinite tower of vector mesons plus pions~\cite{Harada:2006di,Harada:2010cn}. This model is justified in the large $N_c$ and large 't Hooft $\lambda = N_c g^2_{\rm YM}$ limit and the chiral limit.
In these limits, there are only two parameters in the model and they are fixed from meson dynamics. In the "bottom-up" approach, instead of reducing the action from the  gravity sector, it is constructed from QCD in five dimensional anti-de Sitter space~\cite{DaRold:2005zs,Erlich:2005qh} using the duality dictionary. In these models, in addition to the gauge fields, the effect of the explicit breaking of chiral symmetry encoded in the vacuum expectation value (VEV) is included through that of the bulk scalars. In both approaches, the generic model action has the structure
\begin{equation}
S_5 = S^{\rm DBI}_5 + S^{\rm CS}_5,
\label{eq:generic5dS}
\end{equation}
where $S^{\rm DBI}_5$ is the 5D Dirac-Born-Infeld (DBI) part and $S^{\rm CS}_5$ is the 5D Chern-Simons (CS) part. For different models the expressions of $S^{\rm DBI}_5$ could be different. In this review, we will focus on the top-down holographic model. The application of the bottom-up approach to the baryon structure is reviewed by Pomarol and Wulzer in \cite{BReditor}. No application of the {bottom-up } approach to nuclear and dense matter is available in the literature at this moment.

\subsection{The holographic model from the ``top-down" approach and its BPS limit}

Since what we are interested in is the holographic model of QCD, we consider the SS model~\cite{Sakai:2004cn,Sakai:2005yt} which has the properties closest to QCD. In this model, the DBI part is written in terms of the pure burk gauge fields as
\begin{equation}
S^{\rm DBI}_5 \approx S_{\rm YM} ={} -\kappa \int d^4x dz \frac{1}{2e(z)^2}
\, \mbox{tr}\, \mathcal{F}^2
\label{dbiaction}
\end{equation}
with $\kappa=\frac{\lambda N_c}{216\pi^3}$. $e(z)$ is the effective YM coupling that depends
on the holographic direction $z$ and is proportional to the KK mass as $M_{KK}^{-1/2}$. The 5D Chern-Simons term is
\begin{equation}
S^{\rm CS}_5 = \frac{N_c}{24\pi^2} \int d^4x dz \mbox{tr}\,\left(\mathcal{A}\mathcal{F}^2 + \frac{i}{2}\mathcal{A}^3\mathcal{F} - \frac{1}{10}\mathcal{A}^5\right).
\label{csaction}
\end{equation}
We use the index $m=(\mu,z)$ with $\mu=0,1,2,3$.
The gravity enters in the $z$ dependence of the YM coupling, giving rise to the warping of the space. $\mathcal{A} = \mathcal{A}_\mu dx^\mu + \mathcal{A}_z dz$ is the five-dimensional
U($N_f$) gauge field and $\mathcal{F} = d \mathcal{A} + i \mathcal{A} \mathcal{A}$ is its field strength. In the case of $N_f=2$ flavors,
\begin{equation}
\mathcal{A}=A_{\rm SU(2)}+\frac{1}{2} \tilde{A}_{\rm U(1)}.
\end{equation}
The resulting YM action is
\begin{equation}
S_{\rm YM}=-\kappa\int d^4x dz\frac{1}{2e^2(z)} \left( \,\mbox{tr}\, F_{mn}^2 +
\frac 12 \tilde{F}_{mn}^2 \right),
\label{ym}
\end{equation}
and the CS term
\begin{equation}
S_{\rm CS}=\frac{N_c}{16\pi^2} \int\tilde{A} \wedge \mbox{tr} F^2 +
\frac{N_c}{96\pi^2}\int {\tilde A}\wedge {\tilde{F}}^2.
\label{CS}
\end{equation}
In Eqs.~(\ref{ym}) and (\ref{CS}) $F_{mn}$ is the field strength for the SU(2) gauge field
and $\tilde{F}_{mn}$ stands for the field strength of the U(1) gauge field.

We should stress that the SS model -- which is holographically dual to QCD in the large
$N_c$ and $\lambda$ limit (and the chiral limit) -- has no free parameters. To leading order in $\lambda$, that is, to $O(\lambda)$, $e(z)$ is a constant, so the 5D YM action can be taken to be in flat space. The Chern-Simons term, which comes at $O(N_c\lambda^0)$, does not contribute. Thus to leading order, i.e., $O(N_c\lambda)$, the static baryon -- which is $B=1$ is given by the instanton solution that is self-dual~\cite{Kaplunovsky:2012gb}.

Let us first look at the instanton given by the SS action in the leading order, say, $O(N_c\lambda)$.  At this order only  the $S_5^{\rm DBI}$ in flat space contributes and the $U(1)$ degrees of freedom decouple from $SU(2)$ degrees of freedom. The resulting 5D Yang-Mills theory, in unit of an arbitrary mass dimension, is
\begin{equation}
S = {} -\frac 12\int \mbox{tr} F_{mn}^2 d^4x\, dz ,
\label{BPSaction}
\end{equation}
where
\begin{equation}
F_{mn} = \partial_m A_n-\partial_n A_m +[A_m,A_n]
\end{equation}
with $A_m = T^a A^a_m$ normalized as $\mbox{tr}(T^a T^b) = \frac 12 \delta_{ab}$.
The gauge field transforms
\begin{equation}
A_m \rightarrow g (A_m+\partial_m) g^{-1}
\end{equation}
with $g\in SU(2)$.

The static energy coming from the action (\ref{BPSaction}), known as BPS action,
has a well-known bound, the Bogomol'nyi bound,
\begin{equation}
E \geq 8\pi^2 B ,
\label{bound}
\end{equation}
where $B$ is the instanton number representing baryon number
\begin{equation}
B = \frac{1}{16\pi^2} \int \mbox{tr} (F_{MN} {}^*F_{MN}) d^3 x \, dz ,
\label{B}
\end{equation}
in which $M=1,2,3,z$ and ${}^*F_{MN}=\frac 12\epsilon_{MNAB}^{} F_{AB}^{}$ is the dual field strength.
Now the bound is satisfied if $F_{MN}$ is self-dual, i.e.,
\begin{equation}
F_{MN} = {}^*F_{MN}.
\label{selfdual}
\end{equation}
This means that the energy of the system cannot be lower than the bound. In other words, a system of instanton number $A$ describing mass-number $A$ nucleus has the binding energy equal to zero. This can be referred to as ``BPS skyrmion" as will be explained below.

\subsection{Packing vector mesons in an infinite tower}

The 5D action (\ref{dbiaction}), when KK-reduced to 4D,  can be written as the sum of an infinite number of  vector and axial -vector mesons possessing hidden local symmetry~\cite{Sakai:2005yt}. In leading order, the gauge coupling constant will be a constant independent of metric warping with no coupling to the infinite tower of iso-scalar vector mesons. The skyrmion constructed with this infinite tour will then be equivalent to the instanton discussed above.  A highly illustrative study was made by Sutcliffe~\cite{Sutcliffe:2010et,Sutcliffe:2011ig} which showed how the vector mesons entered in the tower.

Start with the pion only with all the vector mesons set equal to zero.  This is the Skyrme model. The soliton static energy is $\sim 1.24$ times the Bogomol'nyi bound (\ref{bound})\footnote{This corresponds in the case of the Skyrme Lagrangian to the Faddeev bound $12\pi^2 B$.}
\begin{equation}
E^{(0)}= 1.235\, (8\pi^2 B).
\label{1skyrmion}
\end{equation}
Next when the lowest-lying vector meson $\rho$ is included, the soliton mass is drastically reduced  to
\begin{equation}
E^{(1)}=1.071(8\pi^2 B).
\end{equation}
The next-lying axial-vector meson $a_1^{}$ brings this further down to
\begin{equation}
E^{(2)}=1.048(8\pi^2 B).
\end{equation}
Since the full tower of iso-vector vector mesons brings this to the bound $E^{(\infty)}=8\pi^2B$,
it follows that the high-lying vector mesons make the theory flow to a conformal theory.
That the lowest-lying vector meson $\rho$ does nearly all the work in flowing to the conformality is
reminiscent of the near complete saturation of the charge sum rule
of the pion~\cite{Sakai:2004cn,Sakai:2005yt} and nucleon~\cite{Hong:2007dq,Hong:2007ay,Hong:2007kx,Hashimoto:2008zw} form factors.

\begin{figure}[H]
\begin{center}
\includegraphics[width=7cm]{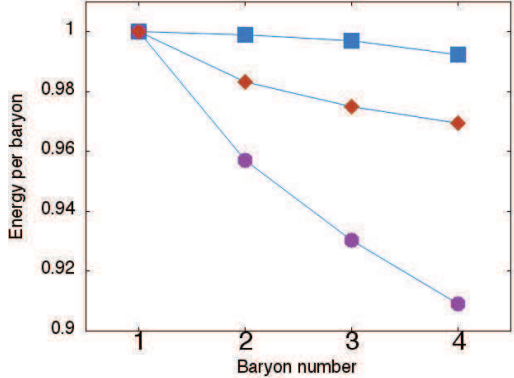}
\caption{
  The energy per baryon, in units of the single baryon energy vs. baryon number up to $A=4$.
  Squares are the experimental data, circles are
  the skyrmion energies with only pions, and
  diamonds are the skyrmion energies in the extended theory with
  pions, $\rho$ and $a_1$ mesons. This figure is taken from~\cite{Sutcliffe:2011ig}.
}\label{sutcliffe}
\end{center}
\end{figure}

Now suppose that the same packing strategy is applied to many-skyrmion or nuclear systems. The above result implies that  the nuclear binding energy, $\sim 24\%$ of the nucleon mass with the pion field only, will reduce to near zero as the higher-lying vector mesons are packed in. This tendency is beautifully illustrated in Fig.~\ref{sutcliffe}.

\subsection{The $\omega$ meson}

In the SS holographic QCD  model, the $\omega$ meson figures at the next order in $\lambda$.  It figures via the Chern-Simons term.  In standard paradigm of nuclear physics, as stressed above, the $\omega$ degree of freedom is indispensable. In chiral perturbation theory, its effect, at least partially, is captured in short-range three-body and higher-body forces brought in by high-order derivative counter terms in the chiral Lagrangian. It is also indicated in lattice simulations of nuclear forces~\cite{Aoki:2012th}. To confront the predictions of the holographic model on nuclear physics with the empirical values, therefore, one should consider the warped space in the YM action and  the Chern-Simons term. By dimensionally de-constructing the SS model and keeping only the lowest-lying vector mesons $\rho$ and $\omega$, the $\omega$ meson is indeed found to be significant for describing the nuclear force~\cite{Ma:2012kb,Ma:2012zm,Ma:2013zpa,Ma:2013ooa,Ma:2013ela,Ma:2013vga}.

Here we analyze how the $\omega$ influences the flow to conformality seen above with the iso-vector vector mesons.
To explore the effect of the $U(1)$ degrees of freedom, one should include the effect of the CS term. The CS term is background-independent and hence should be independent of the warping. Using our energy unit, we have the BPS mass $M_{\rm BPS} \approx \frac{\lambda N_c}{27\pi} M_{KK}
\approx 559$~MeV~\cite{Hong:2007dq,Hong:2007ay,Hong:2007kx}%
\footnote{Here we used $\lambda=16.66, M_{KK} =948$~MeV determined from our inputs.},
in agreement with Ref.~\cite{Sutcliffe:2010et}.
When the CS term contribution is added, we get $M_{\rm BPS-CS} = M_{\rm BPS} + \sqrt{\frac{2}{15}} N_c M_{KK}
\approx 1038$~MeV which shows that the contribution from CS term to the soliton mass is significant.
Along the same procedure applied before, we found that, for the given $\mathcal{M}$, using the HLS given by Eq.~\eqref{eq:Lag_HLS} in the following and determining the low energy constants by the 5D YM theory~\eqref{BPSaction}, the results are~\cite{Ma:2012kb,Ma:2012zm}:
\begin{itemize}
\item $\mathcal{M}=\pi, \rho,\omega$:
\begin{eqnarray}
M_{\rm ``BPS"}(\pi,\rho,\omega)&\approx & 1162\ {\rm MeV} .
\label{eq:MassBPSPRW}
\end{eqnarray}
\item $\mathcal{M}=\pi, \rho$:
\begin{eqnarray}
M_{\rm ``BPS"}(\pi,\rho)&\approx & 577\ {\rm MeV}.
\label{eq:MassBPSPR}
\end{eqnarray}
\item $\mathcal{M}=\pi$:
\begin{eqnarray}
M_{\rm ``BPS"}(\pi)&\approx & 673\ {\rm MeV}.
\label{eq:MassBPSP}
\end{eqnarray}
\end{itemize}
The $\omega$ meson is seen to block the flow to the conformal fixed point. In Eqs.~\eqref{eq:MassBPSPRW},~\eqref{eq:MassBPSPR} and ~\eqref{eq:MassBPSP}, the subindex ``BPS" means the parameters are calculated from the 5D YM theory~\eqref{BPSaction}.

We next examine the effect of the tower of the infinite vector mesons from the SS model in which the warp effect is included. A direct calculation of the mass of the topological object, here, the {\it approximate} instanton, yields the mass~\cite{Hong:2007dq,Hong:2007ay,Hong:2007kx,Hata:2007mb,Hashimoto:2008zw}
\begin{equation}
M_{\rm (approx)instanton}\simeq 1800\  \mbox{MeV},
\label{infinite1}
\end{equation}
with $f_\pi=92.4$~MeV and $\lambda=17$ fixed in the meson sector~\cite{Sakai:2004cn,Sakai:2005yt}. Note that the mass~\eqref{infinite1} corresponds to the mass of a skyrmion with the effect of the infinite tower of vector mesons in a warped space and the Chern-Simons term taken into account.

By dimensionally de-constructing the 5D SS model to the 4D hidden local symmetric mesonic theory, one can check the effect of the tower of vector mesons. It was found that~\cite{Ma:2012kb,Ma:2012zm}:
\begin{itemize}
\item $\mathcal{M}=\pi, \rho,\omega$:
\begin{eqnarray}
M_{\rm SS}(\pi,\rho,\omega)&\approx & 1184\ {\rm MeV} .
\label{eq:MassHLSPRW}
\end{eqnarray}
\item $\mathcal{M}=\pi, \rho$:
\begin{eqnarray}
M_{\rm SS}(\pi,\rho)&\approx & 835\ {\rm MeV} .
\label{eq:MassHLSPR}
\end{eqnarray}
\item $\mathcal{M}=\pi$:
\begin{eqnarray}
M_{\rm SS}(\pi)&\approx & 922\ {\rm MeV} .
\label{eq:MassHLSP}
\end{eqnarray}
\end{itemize}
In the above calculation, the parameters are fixed by the two  empirical values, $f_\pi$ and $m_\rho$,  in the meson sector~\cite{Sakai:2004cn,Sakai:2005yt}. It was found that, if the parameter $M_{KK}$ is reduced to $\sim 500$~MeV, one  gets  $\sim 950$~MeV for the soliton mass and $\sim 300$~MeV for the $\Delta$-$N$ mass difference, both consistent with experiments~\cite{Hata:2007mb}. {In addition, by using the truncated model $\mathcal{M}=\pi, \rho,\omega$, one can fit the nucleon mass and the $\Delta-N$ mass splitting if one uses $f_\pi \simeq 62.2$~MeV and $m_\rho \simeq 417.5$~MeV~\cite{Harada:2012dm}.} In sum, what is found is the following: In the holographic QCD of SS, to the leading $\lambda$ order, the more iso-vector vector mesons are included, the lighter soliton mass becomes and the sum of the contributions from the infinite tower reduces the soliton mass to the BPS instanton limit. The residual interaction between the skyrmions (as quasiparticles) gets weaker and the size becomes smaller, all going in the right direction. However this tendency gets blocked at the next order in $\lambda$, namely at $O(\lambda^0)$, primarily by the presence of the $\omega$ meson present in the CS term with the effect of metric warping less significant. This is at odds with nature. We address this problem and point at a possible resolution with intervention of a scalar dilaton in the effective Lagrangian.

\section{The Effect of the Lowest-Lying Vector Mesons $\rho$ and $\omega$ and the Scalar Dilaton $\sigma$}

\label{sec:vector}

In the above section, we have seen that the infinite tower of the vector meson resonances plays an important role in the skyrmion physics. Here we focus on the nuclear matter from the chiral effective theory including the lowest-lying vector resonances, $\rho$ and $\omega$, based on the HLS approach.

The effect of the lowest-lying vector mesons on dense skyrmion matter was first studied in Ref.~\cite{Park:2003sd} by using a minimal model including vector mesons~\cite{Meissner:1999pe}. In this model, the $\omega$ meson couples only to the baryon density through $\omega_\mu B^\mu$ (with $B^\mu$ being the baryon number current) representing the homogeneous Wess-Zumino term present in HLS theory.  From this model, the attractive force due to the $\rho$ meson and repulsive force arising from $\omega$ meson between nucleons in both skyrmion and skyrmion matter were illustrated clearly. More recently,  this study was refined in HLS  including the next-to-leading order terms of the chiral counting~\cite{Ma:2012kb,Ma:2012zm,Ma:2013zpa,Ma:2013ooa,Ma:2013ela,Ma:2013vga}. In these calculations, the anomalous part of the effective theory that encodes the $\omega$  contribution corresponding to the $U(1)$ gauge field in the CS term of the SS model was fully taken into account.

\subsection{Hidden local symmetry Lagrangian to $O(p^4)$}

As will be mentioned below, there is an indication that HLS to the leading order, that is, $O(p^2)$, has a hidden power that is not obvious from general EFT considerations. We comment on this later. Here we will consider up  to $O(p^4)$ in the power expansion.

The most general HLS Lagrangian responsible for the soliton mass -- which is $O(N_c)$ --  can be expressed as
\begin{eqnarray}
\mathcal{L}_{\rm HLS} & = &
\mathcal{L}_{(2)} + \mathcal{L}_{(4)} + \mathcal{L}_{\rm anom} ,
\label{eq:Lag_HLS}
\end{eqnarray}
with
\begin{eqnarray}
\mathcal{L}_{(2)} & = &
f_\pi^2 \,\mbox{Tr}\, \left( \hat{\alpha}_{\perp\mu} \hat{\alpha}_{\perp}^{\mu} \right)
+ a f_\pi^2 \,\mbox{Tr}\, \left(\hat{\alpha}_{\parallel\mu} \hat{\alpha}_{\parallel}^{\mu} \right) + \mathcal{L}_{\rm kin},
\label{eq:hlsp2}
\end{eqnarray}
where $f_\pi$ is the pion decay constant, $a$ is the parameter of the HLS. The two 1-forms,  $\hat{\alpha}_{\parallel \mu}$ and $\hat{\alpha}_{\perp \mu}$ in \eqref{eq:hlsp2} are defined by
\begin{eqnarray}
& & \hat{\alpha}_{\parallel \mu} =
\frac{1}{2i} (D_\mu \xi_R^{} \cdot \xi_R^\dagger + D_\mu \xi_L^{}\cdot \xi_L^\dagger), \label{eq:defalphapara}\\
& & \hat{\alpha}_{\perp \mu} =
\frac{1}{2i} (D_\mu \xi_R^{}\cdot \xi_R^\dagger - D_\mu \xi_L^{}\cdot \xi_L^\dagger),\label{eq:defalphaperp}
\end{eqnarray}
with the chiral fields $\xi_L^{}$ and $\xi_R^{}$, which in the unitary gauge are
\begin{eqnarray}
\xi_L^\dagger & = & \xi_R^{} = e^{i \pi/2f_\pi} \equiv \xi
\qquad \mbox{with} \qquad \pi = \bm{\pi} \cdot \bm{\tau},
\end{eqnarray}
where $\bm{\tau}$'s are the Pauli matrices.
The covariant derivative associated with the HLS is defined as
\begin{eqnarray}
D_\mu \xi_{R,L} = (\partial_\mu - i V_\mu) \xi_{R,L}^{}
\end{eqnarray}
where $V_\mu$ represents the gauge boson of the HLS,
\begin{eqnarray}
V_\mu = \frac{1}{2} \left( g_\omega\omega_\mu + g_\rho \rho_\mu \right)
\end{eqnarray}
and
\begin{eqnarray}
\rho_\mu = \bm{\rho}_\mu \cdot \bm{\tau} =
\left( \begin{array}{cc}
  \rho_\mu^0  & \sqrt{2} \rho_\mu^+ \\
  \sqrt{2} \rho_\mu^- &  -\rho_\mu^0
\end{array} \right) .
\end{eqnarray}
In \eqref{eq:hlsp2}, $\mathcal{L}_{\rm kin}$ is the kinetic term of vector mesons with
\begin{eqnarray}
\mathcal{L}_{\rm kin} ={}
- \frac{1}{2g_\rho^2} \mbox{Tr}\, \left( V_{\mu\nu}^{(\rho)} V^{(\rho), \mu\nu} \right) - \frac{1}{2g_\omega^2} \mbox{Tr}\, \left( V_{\mu\nu}^{(\omega)} V^{(\omega), \mu\nu} \right), \label{eq:hlskin}
\end{eqnarray}
with the field-strength tensors of the vector mesons
\begin{eqnarray}
V_{\mu\nu}^{(\rho)} & = & \partial_\mu \left(\frac{1}{2}g_\rho\rho_\nu\right) - \partial_\nu \left(\frac{1}{2}g_\rho\rho_\mu\right) - i \left[\left(\frac{1}{2}g_\rho\rho_\mu\right),\left(\frac{1}{2}g_\rho\rho_\nu\right)\right] , \nonumber\\
V_{\mu\nu}^{(\omega)} & = & \partial_\mu \left(\frac{1}{2}g_\omega\omega_\nu\right) - \partial_\nu \left(\frac{1}{2}g_\omega\omega_\mu\right).
\end{eqnarray}
The $O(p^4)$ Lagrangian in Eq.~\eqref{eq:Lag_HLS} is given by
\begin{eqnarray}
\mathcal{L}_{(4)} & = &
y_1^{} \mbox{Tr} \Bigl[ \hat{\alpha}_{\perp\mu}^{} \hat{\alpha}_\perp^\mu
\hat{\alpha}_{\perp\nu}^{} \hat{\alpha}_\perp^\nu \Bigr]
+ y_2^{} \mbox{Tr} \Bigl[ \hat{\alpha}_{\perp\mu}^{} \hat{\alpha}_{\perp\nu}^{}
\hat{\alpha}^\mu_\perp \hat{\alpha}^\nu_\perp \Bigr]\nonumber \\
& &{}
+ y_3^{} \mbox{Tr}
\left[ \hat{\alpha}_{\parallel\mu}^{} \hat{\alpha}_\parallel^\mu
\hat{\alpha}_{\parallel\nu}^{} \hat{\alpha}_\parallel^\nu \right]
+ y_4^{} \mbox{Tr}
\left[ \hat{\alpha}_{\parallel\mu}^{} \hat{\alpha}_{\parallel\nu}^{}
\hat{\alpha}^\mu_\parallel \hat{\alpha}^\nu_\parallel \right]
\nonumber \\
& &{}
+ y_5^{} \mbox{Tr}
\left[ \hat{\alpha}_{\perp\mu}^{} \hat{\alpha}_\perp^\mu
\hat{\alpha}_{\parallel\nu}^{} \hat{\alpha}_\parallel^\nu \right]
+ y_6^{} \mbox{Tr}
\left[ \hat{\alpha}_{\perp\mu}^{} \hat{\alpha}_{\perp\nu}^{}
\hat{\alpha}^\mu_\parallel \hat{\alpha}^\nu_\parallel \right]\nonumber \\
& &{}
+ y_7^{} \mbox{Tr}
\left[ \hat{\alpha}_{\perp\mu}^{} \hat{\alpha}_{\perp\nu}^{}
\hat{\alpha}^\nu_\parallel \hat{\alpha}^\mu_\parallel \right]
\nonumber \\ && \mbox{}
+ y_8^{} \left\{
\mbox{Tr} \left[ \hat{\alpha}_{\perp\mu}^{} \hat{\alpha}_\parallel^\mu
\hat{\alpha}_{\perp\nu}^{} \hat{\alpha}_\parallel^\nu \right]
+ \mbox{Tr} \left[ \hat{\alpha}_{\perp\mu}^{} \hat{\alpha}_{\parallel\nu}^{}
\hat{\alpha}_\perp^\nu \hat{\alpha}_\parallel^\mu \right] \right\}\nonumber \\
& &{}
+ y_9^{} \mbox{Tr}
\left[ \hat{\alpha}_{\perp\mu}^{} \hat{\alpha}_{\parallel\nu}^{}
\hat{\alpha}^\mu_\perp \hat{\alpha}^\nu_\parallel \right],
\nonumber\\
 &  & +
i z_4^{} \mbox{Tr}
\Bigl[ V_{\mu\nu}^{(\rho)} \hat{\alpha}_\perp^\mu \hat{\alpha}_\perp^\nu \Bigr]
+ i z_5^{} \mbox{Tr}
\left[ V_{\mu\nu}^{(\rho)} \hat{\alpha}_\parallel^\mu \hat{\alpha}_\parallel^\nu \right].
\end{eqnarray}
Note that $V_{\mu\nu}^{(\omega)}$ does not appear in the $z_4$ and $z_5$ terms.
Finally, the anomalous parity the hWZ term, $\mathcal{L}_{\rm anom}$, is
\begin{eqnarray}
\Gamma_{\rm hWZ} & = & \int d^4x \mathcal{L}_{\rm anom} = \frac{N_c}{16\pi^2}\int_{M^4}
\sum_{i=1}^3 c_i \mathcal{L}_i ,\label{hWZ}
\end{eqnarray}
where $M^4$ stands for the 4-dimensional Minkowski space and
\begin{subequations}
\begin{eqnarray}
\mathcal{L}_1 & = & i \, \mbox{Tr}\,
\bigl[ \hat{\alpha}_{\rm L}^3 \hat{\alpha}_{\rm R}^{}
 - \hat{\alpha}_{\rm R}^3 \hat{\alpha}_{\rm L}^{} \bigr], \\
\mathcal{L}_2 & = & i \, \mbox{Tr}\,
\bigl[ \hat{\alpha}_{\rm L}^{} \hat{\alpha}_{\rm R}^{}
\hat{\alpha}_{\rm L}^{} \hat{\alpha}_{\rm R}^{} \bigr]  ,  \\
\mathcal{L}_3 & = & \mbox{Tr}\,
\bigl[ F_{\rm V} \left( \hat{\alpha}_{\rm L}^{} \hat{\alpha}_{\rm R}^{}
 - \hat{\alpha}_{\rm R}^{} \hat{\alpha}_{\rm L}^{} \right) \bigr] ,
\end{eqnarray}
\end{subequations}
in the 1-form and 2-form notations with
\begin{eqnarray}
\hat{\alpha}_{L}^{} & = & \hat{\alpha}_\parallel^{} - \hat{\alpha}_\perp^{},
\nonumber \\
\hat{\alpha}_{R}^{} & = & \hat{\alpha}_\parallel^{} + \hat{\alpha}_\perp^{},
\nonumber \\
F_V & = & dV - i V^2.
\end{eqnarray}

In the Lagrangian (\ref{eq:Lag_HLS}) there appear numerous many parameters,  $f_\pi$, $a$, $g_\rho$, $g_\omega$, $y_i (i=1,\cdots,9)$, $z_i(i=4,5)$, and $c_i(i=1,2,3)$. It is very difficult to fix them all in a phenomenological way. A possible approximation is to use a recently developed holographic QCD model in which the coefficients can be completely fixed by means of a set of  ``master formulae"  that  match the four-dimensional effective (HLS) theory to the five dimensional holographic QCD (hQCD) model. In the large $N_c$ and large $\lambda$ limit, the hQCD has, as mentioned above, two parameters which can be related to the empirical values of the pion decay constant and the $\rho$ meson mass. Here we take the SS model~\cite{Sakai:2004cn,Sakai:2005yt} and the following empirical values:
\begin{eqnarray}
f_\pi = 92.4 \mbox{ MeV}, \quad m_\omega = m_\rho = 775.5 \mbox{ MeV}.
\label{eq:PMS}
\end{eqnarray}

\subsection{Skyrmion properties in hidden local symmetry}

In addition to the the soliton masses obtained above -- in Eqs.~\eqref{eq:MassHLSPRW},~\eqref{eq:MassHLSPR} and~\eqref{eq:MassHLSP}, one can also study other properties, i.e.,  the soliton winding number radius $\sqrt{\langle r^2\rangle_W}$ and energy radius $\sqrt{\langle r^2\rangle_E}$. The results are:
\begin{itemize}
\item $\mathcal{M}=\pi, \rho,\omega$:
\begin{eqnarray}
\sqrt{\langle r^2\rangle_W} &\approx & 0.433\ {\rm fm}, \nonumber \\
\sqrt{\langle r^2\rangle_E} &\approx& 0.608\ {\rm fm}.
\label{eq:rHLSPRW}
\end{eqnarray}
\item $\mathcal{M}=\pi, \rho$:
\begin{eqnarray}
\sqrt{\langle r^2\rangle_W} &\approx & 0.247 \ {\rm fm}, \nonumber \\
\sqrt{\langle r^2\rangle_E} &\approx& 0.371 \ {\rm fm}.
\label{eq:rHLSPR}
\end{eqnarray}
\item $\mathcal{M}=\pi$:
\begin{eqnarray}
\sqrt{\langle r^2\rangle_W} &\approx & 0.309 \ {\rm fm}, \nonumber \\
\sqrt{\langle r^2\rangle_E} &\approx& 0.417\ {\rm fm}.
\label{eq:rHLSP}
\end{eqnarray}
\end{itemize}
These results support the intuitive picture that the soliton is compressed due to the attractive force from the $\rho$ meson, while it is expanded by the repulsive force from the $\omega$ meson. The potentially important contribution from a dilaton, particularly to nuclear and dense matter, will also figure in the balance between the attraction and the repulsion.

\begin{figure}[H]\centering
\includegraphics[scale=0.3]{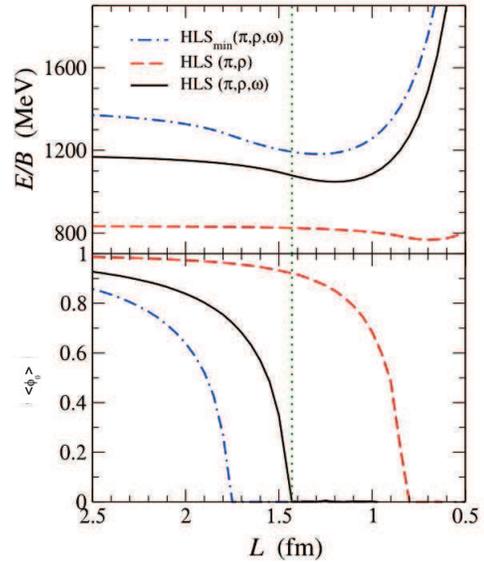}
\caption{The energy per skyrmion and the vacuum expectation value of $\phi_0$ as a function of the crystal size. The vertical line indicates the crystal size corresponding to the normal nuclear density. }
\label{fig:PerSolitonMassHLS}
\end{figure}

Putting the skyrmions given by \eqref{eq:Lag_HLS} onto crystal lattice -- here FCC, one can also study the effect of the lowest-lying vector mesons in  dense skyrmion matter. Here and in what follows, HLS$_{\rm min}(\pi,\rho,\omega)$ will stand for the minimal model of Ref.~\cite{Meissner:1999pe} -- which can be obtained from~\eqref{eq:Lag_HLS} by {switching off } ${\cal L}_{(4)}$ and taking $c_1 = - c_2 = 2/3$ and $c_3 = 0$ in the ${\cal L}_{\rm anom}$ term. The energy per skyrmion and the vacuum expectation value $\la\phi_0 \ra\equiv \la\bar{q}q\ra$  as functions of the crystal size (density) are plotted in Fig.~\ref{fig:PerSolitonMassHLS}.
From Fig.~\ref{fig:PerSolitonMassHLS} we can draw the following conclusions:  the critical density $n_{1/2}$ at which the half-skyrmion phase appears in ${\rm HLS}(\pi,\rho,\omega)$ is larger than that in ${\rm HLS}_{\rm min}(\pi,\rho,\omega)$. This is because in ${\rm HLS}(\pi,\rho,\omega)$ the skyrmion size is smaller so that it needs a smaller distance to have a significant force between skyrmions. This conclusion is supported by the $n_{1/2}$ in ${\rm HLS}(\pi,\rho)$ which yields a much smaller skyrmion size. From ${\rm HLS}(\pi,\rho)$ one sees that the binding energy in this case is much smaller than that in the other two cases. This is because, the model ${\rm HLS}(\pi,\rho)$ includes only the lowest-lying isovector vector mesons which do the most efficient drive toward the conformal limit. Note also that accounting for the full hWZ terms is important for the location of $n_{1/2}$.

\subsection{Skyrmion matter in hidden local symmetry}

For confronting nature, the relevant observables are the in-medium properties discussed in Sec.~\ref{sec:power}. The interesting quantities are the in-medium pion decay constant $f_\pi^*$ and the effective nucleon mass $m_N^*$. What turns out to be the most remarkable is the close relation between the two in the half-skyrmion phase.  The relation remains valid even in the presence of scale invariance that will be taken into account below.

\begin{figure}[H]\centering
\includegraphics[scale=0.43]{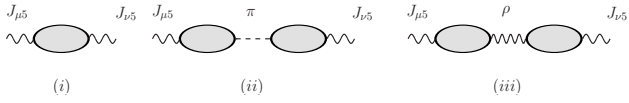}
\caption{Three types of contributions to the correlator of
Eq.~\eqref{eq:defaacorr}: (i) the contact diagram, (ii) the pion exchange diagram,
and (iii) the rho exchange diagram. Shaded blobs stand for the
skyrmion matter interaction vertices. }
\label{fig:AACHLS}
\end{figure}

\begin{figure}[H]\centering
\includegraphics[scale=0.25]{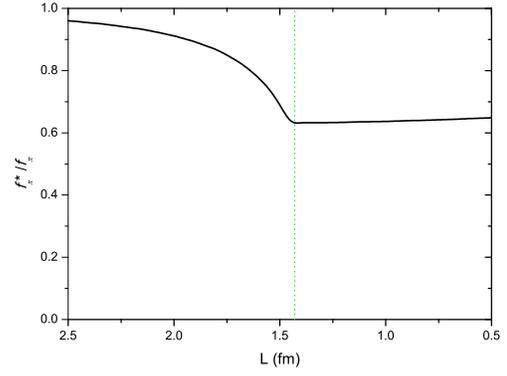}
\caption{Crystal size dependence of the $f_\pi^\ast$ normalized by its vacuum values. }
\label{fig:fpimrhoHLS}
\end{figure}

Consider the pion decay constant. The axial correlator ~\eqref{eq:defaacorr} receives an additional contribution to Fig.~\ref{fig:Correlator} for the Skyrme model from the $\rho$ exchange as depicted in Fig.~\ref{fig:AACHLS}.
The three graphs contribute
\begin{eqnarray}
(i): & if_\pi^2 g_{\mu\nu}\delta^{ab}\left\langle 1 + \frac{1-a}{2}\left[\left(1 - \frac{2}{3}\bm{\phi}_\pi^2\right)-1\right]\right\rangle\nonumber\\
(ii): & {}- if_\pi^2 \frac{p_\mu p_\nu}{p^2}\delta^{ab}\left\langle \left(1 - \frac{2}{3}\bm{\phi}_\pi^2\right)-1\right\rangle \nonumber\\
(iii): & i\delta^{ab}\left\langle \frac{a^2 g^2 f_\pi^4}{p^2 - m_\rho^2}\left(g_{\mu\nu}-\frac{p_\mu p_\nu}{m_\rho^2}\right)\left[\left(1 - \frac{2}{3}\bm{\phi}_\pi^2\right)-1\right]\right\rangle .
\end{eqnarray}
To the order of expansion considered, the sum of the three terms is Lorentz invariant and has the same expression of $f_\pi^\ast$ as that from the Skyrme model, Eq.~\eqref{eq:mediumfpi}. The crystal size dependence of $f_\pi^\ast$ from HLS skyrmion is plotted in Fig.~\ref{fig:fpimrhoHLS}. In the skyrmion phase $n<n_{1/2}$, $f_\pi^\ast$ decreases as density increases. This tendency is supported in observations with deeply bound pionic nuclei. What is noteworthy and is of importance in the development  is that this tendency stops at $n_{1/2}$. After $n_{1/2}$, $f_\pi^\ast$ stays more or less constant nearly (modulo approximations) independent of density. The behavior of  $f_\pi^\ast$ in HLS theory is the same as that in the Skyrme model shown in Fig.~\ref{fig:inmediumpion}. {\it This robust property of the pion decay constant, strikingly independent of the degrees of freedom included, is the key feature of the approach anchored on the topology change. The topology is dictated by the pion field and the $\rho$ field does not carry any influence on the topology.}

\begin{figure}[H]\centering
\includegraphics[scale=0.25]{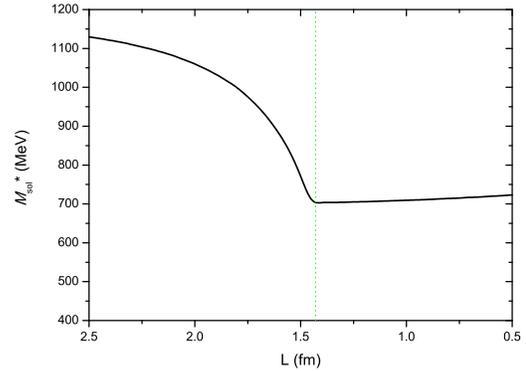}
\caption{Crystal size dependence of the medium modified nucleon mass. }
\label{fig:mNHLS}
\end{figure}

In the present approach, the skyrmion properties  are calculated with the parameters fixed by meson dynamics in HLS. Hence, by taking the medium modified parameters as input, one can calculate the medium-modified skyrmion (or baryon) properties. This allows us to calculate the skyrmion-crystal matter to obtain modified baryon properties with $f_\pi^\ast$ plotted in Fig.~\ref{fig:fpimrhoHLS} as input. The in-medium nucleon mass $m_N^*\sim M_{\rm sol}^\ast$ so obtained is plotted in Fig.~\ref{fig:mNHLS}. This result shows that the density dependence of the nucleon mass is surprisingly similar to that of $f_\pi^\ast$. The simple way to understand this close connection is the large $N_c$ property of the skyrmion model which seems to hold for low as well as high density.  The nucleon mass should behave as $\sim \xi f_\pi^\ast$ with $\xi$ scale independent (and density independent) $O(\sqrt{N_c})$ constant.

\subsection{Scale symmetry: Explicit breaking and impact on dense matter}

\label{omegaproblem}

\label{scale-chiral}

What is strikingly clear from the results discussed above is that the presence of the $\omega$ field substantially worsens the comparison of the skyrmion matter with nature. We show here that what's involved is the role of  the scalar degree of freedom~\cite{MRdilaton}.  In what we discussed so far, no scalar degree of freedom has figured explicitly in the hidden local symmetric Lagrangian, either with the lowest-lying vector mesons or with the infinite tower. The standard strategy in nuclear physics based on effective Lagrangians, both nonlinear sigma model and hidden local symmetric model, is that the effect of the iso-scalar scalar excitations are generated at high order of chiral expansions. However in the approach exploiting HLS, it is natural to consider explicitly the scalar excitations of mass $\sim 600$ MeV, given that this mass scale is comparable to that of vector mesons and that tensor forces, essential in nuclear dynamics, involve the $\rho$ meson in addition to the pion and the $\omega$ meson provide the crucial repulsion that stabilize the nuclear matter.  First in 1991~\cite{Brown:1991kk} and since then~\cite{Park:2008zg,Park:2009mg,Lee:2010sw,Ma:2013ela}, the scalar, interpreted as a Nambu-Goldstone boson, say, dilaton $\sigma$, was introduced to describe the properties of light-quark hadrons on dense medium, as a means of simulating a precursor signal for chiral restoration expected at high density. The basic idea there was that chiral symmetry characterized by the quark condensate $\langle\bar{q}q\rangle$ is locked to the dilaton condensate $\langle\chi\rangle$. It has become clear that describing dense baryonic matter in HLS, above the normal nuclear matter density $n_0$, is unrealistic without incorporating the dilaton degree of freedom. This is because there is an intricate interplay between the role of $\omega$ providing repulsion and that of dilaton  providing attraction. Thus much of the difficulty in the single baryon as well as in baryonic matter we observed above when the $\omega$ degree of freedom is incorporated is caused by what we consider as inadequate treatment of scale symmetry broken both explicitly and spontaneously.   This comment applies to both HLS-skyrmion matter (4D) and holographic matter (5D).

One powerful way to implement scale symmetry in HLS approach was recently suggested by Crewther and Tunstall (CT)~\cite{CT}~\footnote{There is an alternative approach that involves an IR fixed point~\cite{Golterman:2016lsd} in which the role of IR fixed point is
given in terms of a “critical” number of flavors/number of colors
$n_f = N^c_f /N_c$ in what corresponds to the Veneziano limit. A comparison of this approach and C-T was given in Ref.~\cite{Li:2016uzn}.}. This approach is anchored on the possible existence in QCD of an infrared fixed point at which the trace anomaly is to vanish in the chiral limit~\footnote{Whether such an IR fixed exists in QCD with the number of flavors $N_F <8$ is not yet settled. This matter is discussed elsewhere in the context of applications in nuclear system and we won't go into it. This issue, clearly important for the fundamental structure of QCD, is however not so crucial in applications to dense matter because our approach does not require the precise nature of QCD in the chiral limit, which is not honored in dense matter.}

We start with the leading order nonlinear sigma model Lagrangian to which scale symmetry is implemented as written by CT. For simplicity in notation, we work in the chiral limit. The CT Lagrangian, in terms of the conformal compensator field $\chi=f_\sigma e^{\sigma/f_\sigma}$, is
\be
{\cal L} &=&{\cal L}_{\rm inv} + {\cal L}_{anom} +V(\chi),\\
{\cal L}_{\rm inv} & = & c_1 {\frac{f_\pi^2}{4} \left(\frac{\chi}{f_\sigma}\right)^2} {\rm Tr} \left(\partial_\mu U\partial^\mu U^\dagger \right)+c_2\frac 12 \partial_\mu\chi\partial^\mu\chi,\label{inv}\\
{\cal L}_{anom} & = & \Bigg\{(1-c_1) {\frac{f_\pi^2}{4} \left(\frac{\chi}{f_\sigma}\right)^2} {\rm Tr} \left(\partial_\mu U\partial^\mu U^\dagger \right)\nonumber\\
&&\;\;\;{} +(1-c_2)\frac 12 \partial_\mu\chi\partial^\mu\chi\Bigg\}\left(\frac{\chi}{f_\sigma}\right)^{\beta^\prime},\\
V(\chi) & = & \left(\frac{\chi}{f_\sigma}\right)^4 \left[c_3+c_4\left(\frac{\chi}{f_\sigma}\right)^{\beta^\prime}\right] .
\ee

Following \cite{CT},  we set,  in the chiral limit, $c_1=c_2=1+O(p^2)$ which can be arrived at by setting the dilaton field equal to zero for processes that do not involve scalar excitations.  The best way to  understand this relation is that there is hidden scale symmetry in Standard Higgs-type Lagrangian that yields both the scale-symmetric form and the nonlinear sigma model form that can be reached when one dials a constant, respectively,  to weak coupling limit and to strong-coupling limit~\cite{yamawaki}. Keeping to $O(p^2)$ in the leading-order Lagrangian, we have
\be
{\cal L}_{\rm inv}= {\frac{f_\pi^2}{4} \left(\frac{\chi}{f_\sigma}\right)^2} {\rm Tr} \left(\partial_\mu U\partial^\mu U^\dagger \right)+\frac 12 \partial_\mu\chi\partial^\mu\chi +O(p^4).
\ee
Hidden-local-symmetrizing this, we have{~\cite{MRdilaton}}
\be
{\cal L}_{sHLS} & = & f_\pi^2 \left(\frac{\chi}{f_\sigma}\right)^2{\rm
Tr}\left[\hat{a}_{\perp\mu}\hat{a}_{\perp}^{\mu}\right] + a f_\pi^2  \left(\frac{\chi}{f_\sigma}\right)^2{\rm
Tr}\left[\hat{a}_{\parallel\mu}\hat{a}_{\parallel}^{\mu}\right]\nonumber\\
& &{} - \frac{1}{2g^2}{\rm Tr}\left[V_{\mu\nu}V^{\mu\nu}\right] + \frac{1}{2} \partial_\mu \chi \partial^\mu \chi +O(p^4).\label{shls}
\ee
The dilaton potential is of course unaffected.

In the presence of vector fields, the anomalous-parity homogeneous Wess-Zumino term  ${\cal L}_{an}$, Eq.~(\ref{hWZ}) needs to be taken into account.  There are three terms. For an approximate calculation we take only one term in the form, $g\omega_\mu B^\mu$ where $B_\mu$ is the baryon current. It is straightforward to treat all three terms involved therein at the cost of more parameters. The $g\omega_\mu B^\mu$ term considered here is $O(p^4)$, higher order than what's treated above. It cannot however be ignored because it is through this term that the $\omega$ field couples to the other degrees of freedom of HLS Lagrangian. Without it, the $\omega$ does not figure in the interactions. By itself, this term is of scale dimension 4, hence scale-invariant. Therefore it has no coupling to the dilaton field $\chi$. This is however not consistent with scale-chiral symmetry \`a la CT. The correct expression should be~\cite{MRdilaton}
\be
{\cal L}_{hWZ\chi} = \left(c_h+(1-c_h)\left(\frac{\chi}{f_\sigma}\right)^{\beta^\prime}\right)  {\cal L}_{\text {hWZ}}.\label{hWZchi}
\ee
Unlike $c_{1,2}$ in (\ref{inv}), there is no reason why $c_h$ should be close to 1. In fact there is an indication that $c_h=1$ at which there is no coupling to $\chi$ is found to be violently at odds with nature.

In \cite{Park:2003sd}, when skyrmions, obtained  with  the Lagrangians (\ref{shls}) and (\ref{hWZchi}) with $c_h=1$ and $V(\chi)$ of the Coleman-Weinberg type, were put on crystal lattice to simulate dense matter, the contribution to the energy of the system from the hWZ term diverged at high density unless the pion decay constant (or equivalently the $\omega$ mass) went to $\infty$. This defect can be eliminated if the explicit breaking of scale symmetry dampens the strength of the hidden gauge coupling $g$. In \cite{Park:2008zg}, the problem was resolved when the factor $(\frac{\chi}{f_\sigma})^{n}$ with $1< n\lsim 3$ was {\it arbitrarily} multiplied to ${\cal L}_{hWZ}$. This happens because in medium $f_\sigma^*/f_\sigma$ decreases monotonically at increasing density, therefore the coupling gets sufficiently suppressed.  This means that the solution to the problem found in \cite{Park:2008zg} corresponds to (\ref{hWZchi}) with{~\cite{MRdilaton}}
\be
1 <\beta^\prime\lsim  3, \ \ c_h\approx 0.
\ee
It seems that this solution is consistent with the bound for the anomalous dimension of $G^2$ $\beta^\prime=\gamma_{F^2,IR}$ given in \cite{shrock}
\be
\gamma_{F^2,IR} \leq 3.
\ee
If this turns out to be a possible solution to the problem, then the analysis reported above would need to be reexamined with the potential $\beta^\prime$ corrections in the hWZ terms including the CS term in holographic models~\cite{Ma:2013ela}.

{\it What we learned in this subsection is that to correctly describe baryonic matter at high density, scalar degrees of freedom are indispensable. They could perhaps be generated at high (chiral) orders in perturbative scheme. In the spirit espoused in this review, they could come in as a dilaton in the broken scale symmetry. For this, it seems essential, if the analysis described in this subsection is correct, that how the scale symmetry is explicitly broken has to be understood. What we have observed above regarding the disastrous role of the $\omega$ meson in dense matter in the presence of a dilaton indicates the necessity of the dilaton coupling to the $\omega$ in the hWZ term in 4D skyrmion matter or the Chern-Simons term in 5D instanton matter. And this requires understanding how the explicit  scale symmetry breaking manifests in dense medium. This is an open problem to be resolved.}

\section{Other Developments}

In this section we discuss a few interesting developments that are not directly related to what we discussed above anchored on topology change present in the description of dense matter with skyrmions on crystal lattice. The premise is that the effective Lagrangian that is taken has a connection to QCD in the sense of Weinberg's ``folk theorem" on effective field theories~\cite{weinberg}. The basic assumption is that the theory has a valid contact with strong interactions at low energy, beginning with low-energy theorems based on current algebras.  Thus the standard nuclear physics approach to dense matter has been to first start with pions only with baryons put in explicitly or brought out as solitons, describe well nuclear matter and then extrapolate to the regime that is outside of the range that is fit to the theory. Here we discuss a few cases where this standard procedure is not adhered to.

\subsection{The (near-)BPS skyrmion model}

\label{sec:BPS}

In the ``derivation" of the small binding energy from a flat-space YM Lagrangian discussed above, the relevant degrees of freedom in nuclear physics, namely, the pion and the lowest-lying vector mesons, are augmented with the infinite tower leading to the BPS structure. Now instead of adding the infinite tower to arrive at the BPS limit, one simply posits a BPS structure in 4D.  So the philosophy would then be, instead of starting from a chiral Lagrangian with pions only and building up the tower to go to heavy nuclei, why not start with the simplest structure that could work for heavy nuclei with nearly vanishing binding energy and then go from there making appropriate corrections? This is the point of  view taken by Adam {\it et al.}~\cite{Adam:2010fg,Adam:2010ds}. (For a recent review, see Ref.~\cite{Adam:2015ele}). The parameters entering in this approach will have no direct connection with QCD proper, in a spirit totally different to what has been resorted to above.

The justifications offered for this drastically unorthodox approach are:
\begin{enumerate}
\item The binding energy is zero at the {\it  leading order} even though highly non-linear strong interactions may be involved. One might think of this as an extreme quasiparticle description with no residual interactions.
\item The model effectively describes a perfect fluid with its energy-momentum tensor in the perfect fluid form, and the static energy functional is  invariant under volume-preserving diffeomorphisms.
\item While the standard quasiparticle description of baryonic matter, namely, Landau Fermi-liquid theory, can be approximately given in the mean field of a relativistic field theory, here the quasiparticle picture is to capture the {\it exact} result of the theory. The mean-field result of the relativistic field theory can be an approximation that is obtained by ``averaging" the BPS theory.
\end {enumerate}

In the  BPS skyrme model of~\cite{Adam:2010fg}, the basic quantity is the $SU(2)$ valued field $U(\mathbf{x},t)$ in which topology is lodged.  The Lagrangian picked is
\begin{eqnarray}
{\cal L}_{\rm BPS} = {} - \lambda^2 \pi^4 \mathcal{B}_\mu \mathcal{B}^\mu - \mu^2 V({\rm tr}(U)),
\label{eq:BPSL}
\end{eqnarray}
where $\lambda$ and $\mu$ are parameters that are not related to QCD paramters. They will be fit to nuclear binding energies that are focused on. The potential $V({\rm tr}(U))$  is a suitably chosen potential involving nonderivative terms. In ~\cite{Adam:2014dqa}, it is taken in the forms
\begin{eqnarray}
& &V({\rm tr}(U)) = {} V_\pi  = \frac{1}{2}{\rm tr}(1 - U), \quad \mbox{ model I};\nonumber\\
& & V({\rm tr}(U)) = {} (V_\pi)^2  ,  \qquad\qquad\quad\; \mbox{ model II}.
\end{eqnarray}
The baryon number $B_\mu$ takes the familiar expression
\begin{eqnarray}
\mathcal{B}_\mu = {} \frac{1}{24\pi^2}\epsilon^{\mu\nu\alpha\beta}{\rm tr}\left(U^\dagger \partial_\nu UU^\dagger \partial_\alpha UU^\dagger \partial_\beta U\right).
\end{eqnarray}

First to see that the BPS skyrmion has zero binding energy, one shows that it satisfies the Bogomoln'yi equation. The static energy given by Eq.~(\ref{eq:BPSL}) has the bound
\be
E & = & \int d^3 x \left( \pi^2 \lambda \mathcal{B}_0 \pm \mu \sqrt{V} \right)^2 \mp 2 \pi^2 \lambda \mu \int d^3x   \mathcal{B}_0  \sqrt{V} \nonumber\\
& \geq &{} \mp 2 \pi^2 \lambda \mu \int d^3x   \mathcal{B}_0  \sqrt{V} \nonumber\\
& = & 2\pi^2 \lambda \mu |B| \left< \sqrt{V}\right>_{\mathbb{S}^3},
\label{bound}
\ee
where $ \left\langle \sqrt{V}\right\rangle_{\mathbb{S}^3} $
is the average value of $\sqrt{V}$ on the target space $S^3$. The last equality follows from the condition that $B>0$.
The bound is saturated by the solutions of the BPS equation of motion from (\ref{eq:BPSL}).
\begin{equation} \label{BPS-eq}
\pi^2 \lambda \mathcal{B}_0 \pm \mu \sqrt{\mathcal{U}}=0.
\end{equation}

By coupling gravity to (\ref{eq:BPSL}), one can write the action of the BPS model in curved space as
\begin{equation}
S_{\rm BPS}= \int d^4 x |g|^\frac{1}{2} \left( -\lambda^2 \pi^4 |g|^{-1} g_{\rho\sigma} {\cal B}^\rho {\cal B}^\sigma - \mu^2 {V} \right) .
\label{eq:BPS-Gravity}
\end{equation}
From this action, one can show that the energy-momentum tensor has the perfect fluid form
\begin{equation} \label{perf-fluid}
T^{\rho \sigma} = (p+\varepsilon )u^\rho u^\sigma - pg^{\rho\sigma}
\end{equation}
with the four-velocity $u^\rho$, energy density $\varepsilon$ and pressure $p$ given by
\begin{eqnarray}
u^\rho & = & {\cal B}^\rho / \sqrt{g_{\sigma \pi} {\cal B}^\sigma {\cal B}^\pi} , \nonumber \\
\varepsilon & = & \lambda^2 \pi^4 |g|^{-1} g_{\rho\sigma} {\cal B}^\rho {\cal B}^\sigma + \mu^2 {V} , \nonumber \\
p & = & \lambda^2 \pi^4 |g|^{-1} g_{\rho\sigma} {\cal B}^\rho {\cal B}^\sigma - \mu^2 {V} .
\end{eqnarray}
Going further, it can be shown that the BPS skyrmion is equivalent to a non-barotropic, relativistic perfect fluid in the Eulerian formulation.

The energy of the BPS system described by the Lagrangian (\ref{eq:BPSL})
\be
E = \int d^3x \left\{ \pi^4 \lambda^2 {\cal B}_0^2 +\mu^2 V({\rm tr} U)\right\}
\ee
 has, as stated above, a Bogomoln'yi bound. In fact it has infinitely many BPS solutions saturating the bound. The topology is lodged in the field $U$ which is of the pion field type as in the usual chiral Lagrangian. In this approach, the pion field is a highly correlated  nuclear collective excitation of pionic quantum number that supports a solitonic structure that has no fluctuation component that enters in the low-energy theorems of strong interactions. This would naturally exclude the possibility that  the pion field that enters in this description be directly related, if any, to the asymptotic pions observed in nature.  Assuming that it is of the soltion structure similar to that of large $N_c$ QCD model, i.e., the Skyrme model, one takes the axial symmetric ansatz $U(r)=e^{i\vec{\tau}\cdot\vec{r}\theta (r)}$. With this it is straightforward to calculate the soliton energy for the BPS
 \be
 E_{\rm sol}=\frac{64\sqrt{2}\pi}{15}\mu\lambda N,
 \ee
 where $N=B$ is the winding number, i.e., baryon number.  Note that the soliton energy per particle then is just a constant $\propto \mu\lambda$ given by the combination of the potential and the constant multiplying the topological term, so the energy of $N$ baryon number system is a multiple of baryon number and hence zero binding energy at the level of the soliton.

\begin{figure}[H]\centering
\includegraphics[scale=0.6]{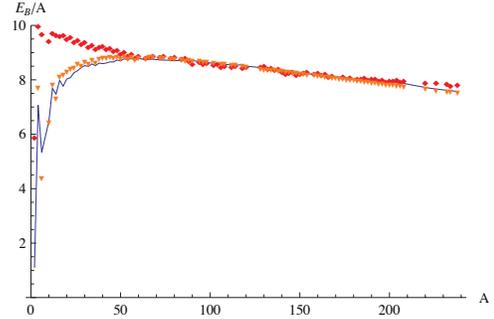}
\caption{Binding energies per nucleon in MeV. The diamond is from the present BPS model. The triangle is from the Weizsacker's formula. And, the solid line is from experimental data. The figure is borrowed from Ref.~\cite{Adam:2013tda}.}
\label{fig:E-bind-BPS}
\end{figure}

 To go beyond the classical soliton structure, one needs to perform the collective quantization. The resulting rotational spectra are given by
 \begin{equation}
E_{rot}  = \frac{\hbar^2}{2} \left( \frac{j(j+1)}{\mathcal J_1} + \frac{i(i+1)}{\mathcal I_1} + \left( \frac{1}{\mathcal I_3}-\frac{1}{\mathcal I_1} - \frac{B^2}{\mathcal J_1} \right) k_3^2 \right),
\end{equation}
 where ${\mathcal J_i}$ and ${\mathcal I_i}$ are, respectively, the angular momentum moment of inertia and the isospin moment of inertia. For even $B$ nuclei, $k_3=0$. The resulting binding energy including the Coulomb energy and isospin violation is given in Fig.~\ref{fig:E-bind-BPS}. The theoretical prediction with only 3 fit-parameters  agrees  very well {with the experimental data and the Weizsacker's empirical formula}.

The model was applied also to nuclear matter and compact-star matter. The latter is done by coupling the BPS skyrmion model to gravity  using the action~\eqref{eq:BPS-Gravity} and the resulting energy-momentum tensor of a perfect fluid \eqref{perf-fluid}.  The maximum mass and  the radius obtained therefrom -- without resorting to the EoS and the Tolman-Oppenheimer-Volkoff (TOV) equation -- come out to be~\cite{Adam:2014dqa}
\begin{itemize}
  \item Model I: \quad\; $M_{\rm max} = 3.734M_\odot$; \;\; $R_{\rm max} = 18.456~$km.
  \item Model II: \quad $M_{\rm max} = 2.439M_\odot$; \;\; $R_{\rm max} = 16.801~$km.
\end{itemize}
\begin{figure}[H]\centering
\includegraphics[scale=0.35]{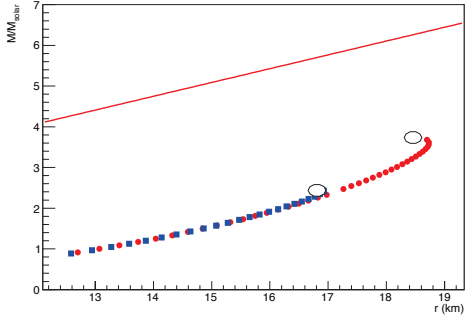}
\caption{Mass-radius relation calculated from the BPS model with potential I (dotted line) and model II (box line). This figure is borrowed from Ref.~\cite{Adam:2014dqa}. }
\label{fig:MRBPS}
\end{figure}
The results plotted in Fig.~\ref{fig:MRBPS} show that  both the mass and the radius -- which depend appreciably on the unknown potential -- tend to be a bit too big in comparison with what's observed~\cite{Demorest:2010bx,Antoniadis:2013pzd}. Given that the model exploits only a few parameters and there is a plenty of room available for refinement, one cannot consider this as a criterion to consider the model failing.

A comment on the contrast of the BPS approach to the current lore of nuclear physics is in order here. There are two key ingredients in the currently predominant attitude among nuclear theorists: The first is that the small binding energy observed in nuclei is an interplay between a big attraction characterized by scalar degrees of freedom and a big repulsion provided by vector degrees of freedom and the second, the theory has be able to accurately satisfy the conditions met both experimentally and theoretically at around nuclear matter density.  The BPS model is contrary to the first in that the BPS ``nucleon" is a topological blob ignorant of the intricate interactions that lead to its property. As it stands, it cannot account for low-energy pionic processes that are described by low-energy theorems that are the ground to effective chiral theories. As for the second, it cannot access the accurately measured nuclear response functions to the {electroweak (EW)} fields. For instance, while it may provide bulk properties of nuclear ground states, it cannot account for the important role of meson-exchange currents that are accurately described by effective chiral Lagrangian approaches. One should however not dismiss this model. It may be possible to start from the BPS matter and develop a theory that can access those processes given by soft-pion theorems. This is in the same spirit as starting from Landau Fermi-liquid fixed point applicable to the regime of nuclear equilibrium density and arriving  at processes involving soft-pions. This is exemplified by relativistic mean-field approaches that are connected to Landau Fermi-liquid theory. There pionic interactions can be -- and are -- readily incorporated.

\subsection{The simplified HLS (SHLS) for dense matter}

\label{subsec:SimplifyHLS}

As mentioned above, the HLS Lagrangian~\eqref{eq:Lag_HLS} yields remarkable predictions for nucleon and nuclear matter. However, even limited to the next-to-the leading order, it involves much too many parameters and while straightforward, is not very illuminating.  Fortunately it turns out that the Lagrangian ~\eqref{eq:Lag_HLS} can be vastly simplified. This is feasible not only in the matter-free vacuum but more significantly in dense matter as described above and elaborated further below.

%%%%%%% Table 1 %%%%%%%%%%%%%%%%%%%%%

\begin{table*}[t]
\caption{\label{table:Ratiohls}
Ratio (\%) of the contribution from each term of HLS to the soliton mass. }
\centering
\begin{tabular}{cccccccccccccccc}
\hline \hline
 $\mathcal{O}(p^2)$ & $y_1$ & $y_2$
& $y_3$ & $y_4$ &  $y_5$
& $y_6$ &  $y_7$ &  $y_8$ &  $y_9$ &  $z_4$ &  $z_5$ &  $c_1$ &  $c_2$ &  $c_3$ \\
\hline
73.89 & 1.29 & 0.35 & 0.03 & ${}-0.02$ & ${}-1.64$ & 0 & ${}-0.07$ & ${}-0.49$ & 3.04 & ${}-13.24$ & 0.22 & 14.21 & 4.54 & 17.88 \\
\hline \hline
\end{tabular}
\end{table*}

%%%%%%%%%%%%%%%%%%%%%%% end table I %%%%%%%%%%%%%%%%%%%%%%%%%%%%%%%%%%%%%%%

To simplify the model , we first analyze how each term of the $O(p^4)$ terms in the HLS sector contributes to  the soliton mass. The results are summarized in Table.~\ref{table:Ratiohls}. From this table,  we can conclude that the $\mathcal{O}(p^2)$, $z_4$  and hWZ terms give dominant contributions, say,  about $97.28\%$ of the total, to the soliton mass. The simplified Lagrangian takes the form
\begin{eqnarray}
{\cal L}_{\rm SHLS} & = & f_\pi^2 {\rm
Tr}[\hat{a}_{\perp\mu}\hat{a}_{\perp}^{\mu}] + a f_\pi^2{\rm
Tr}[\hat{a}_{\parallel\mu}\hat{a}_{\parallel}^{\mu}] -
\frac{1}{2g^2}{\rm Tr}[V_{\mu\nu}V^{\mu\nu}] \nonumber\\
& & {} + i z_4 \,{\rm Tr}\left[ V_{\mu\nu}
\hat{\alpha}_\perp^\mu \hat{\alpha}_\perp^\nu \right] + \cal{L}_{\rm hWZ}.
\label{eq:lagrHSkyr}
\end{eqnarray}

\begin{figure}[H]\centering
\includegraphics[scale=0.4]{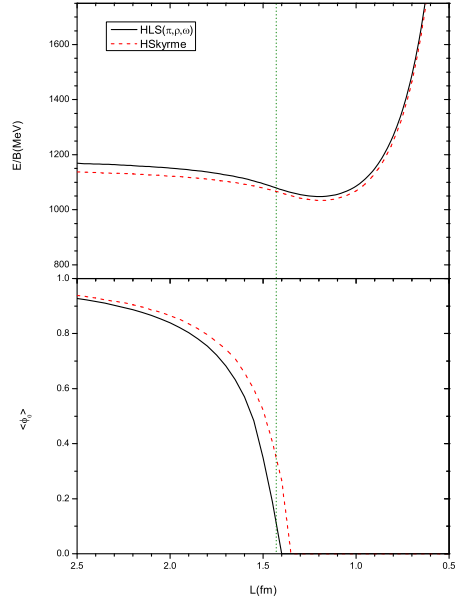}
\caption{Comparision of the $E/B$ and  $\langle \phi_0 \rangle$  calculated from HLS and HSkyrme model as a function of L.}
\label{fig:matterHSkyr}
\end{figure}

We next study the skyrmion matter properties from model~\eqref{eq:lagrHSkyr} and compare the result with that from full HLS~\eqref{eq:Lag_HLS}. Our results of the crystal size $L$ dependence of the per-skymrion energy and $\langle \phi_0 \rangle$ are plotted in Fig.~\ref{fig:matterHSkyr}. This figure tells us that the SHLS model indeed captures the essence of the skyrmion dynamics. Moreover, $n_{1/2}$ is pushed to a somewhat larger value which makes it more consistent with nature.

In summary, the model~\eqref{eq:lagrHSkyr} is verified to capture  the dominant physics of HLS~\eqref{eq:Lag_HLS}. The crucial observation is that, as with the full HLS calculation~\cite{Ma:2013ooa}, the density $n_{1/2}$ at which the transition from skyrmions  to half-skyrmions takes place lies higher than that of normal nuclear matter, $n_0$. The reason that the terms in~\eqref{eq:lagrHSkyr} more or less fully capture  the physics of HLS~\eqref{eq:Lag_HLS} is the following: The $O(p^2)$ terms encode the current algebra of hadron physics, so they are dominant in low-energy nuclear interactions. The $\omega$ meson -- which is indispensable for stability of nuclear matter -- figures through the hWZ terms. Among all the $O(p^4)$ terms -- apart from the  hWZ terms --  the $z_4$ term, involving the strong $\rho$-$\pi$-$\pi$ interaction, contributes most importantly whereas the $z_5$ term, describing $\rho$-$\rho$-$\rho$ interactions, is  suppressed. In fact it is this simplified Lagrangian that through the effect of explicit scale symmetry breaking removes the ``$\omega$ disaster" in dense skyrmion matter discussed in Section \ref{omegaproblem}.

\subsection{Emerging symmetries in  dense matter?}

\label{sec:matter}

That symmetries may emerge in highly correlated systems is rapidly becoming a highly plausible and acceptable concept in physics. This tendency is strikingly visible not only in condensed matter physics but also in particle physics including gravity and dark matter, see e.g., \cite{verlinde}. Dense compact-star matter we have been addressing is equally highly-correlated matter and it would not be an idle conjecture that certain symmetries could emerge as density increases beyond the normal $n_0$. This we claim is indeed the case with both hidden scale symmetry and HLS in dense nuclear systems~\cite{PR-emergence}.  This is a notion eminently novel in nuclear/hadron physics.

An approach that exploits robust topological features of, but bypassing the complexity of, the skyrmion crystal method is a continuum Lagrangian that relies on ``sliding vacuum" for dense matter developed first in 1991~\cite{Brown:1991kk} and then elaborated further in \cite{BR:DD}.  The starting ingredient is an effective Lagrangian constructed along the line of Weinberg's ``folk theorem (FT)" that combines scale symmetry and chiral symmetry (scale-chiral symmetry) of QCD discussed in Section  \ref{scale-chiral}.  The effective Lagrangian is built with the pseudo-NG boson, pion, the lowest-lying vector mesons and the dilaton, encoding hidden scale symmetry and HLS and the baryons as relevant degrees of freedom. The Lagrangian is defined at a scale $\Lambda_M$, referred to as  ``matching scale," at which the correlators of QCD (high-energy scale) and the correlators of the effective theory, dubbed as $bs$HLS (low-energy scale) are optimally matched. It is an open issue as to whether such a matching is feasible. It is usually assumed that the relevant $\Lambda_M$ can be the chiral scale $\sim 4\pi f_\pi\sim 1$ GeV. In practice in nuclear applications, it is taken just above the vector-meson mass scale, $m_V\sim 600-700$ MeV. At the matching, the EFT Lagrangian, $bs$HLS, inherits from QCD the dependence on, among others, nonperturbative QCD quantities, such as the condensates $\la\bar{q}q\ra$, $\la G^2\ra$ etc. injected from QCD correlators into the parameters of the $bs$HLS. Therefore the effective Lagrangian will carry information on the ``vacuum." Now the basic assumption in the present approach is that those condensates scale as the density of the system described by the Lagrangian. This will then constitute a Lagrangian defined in a ``sliding vacuum" which would provide the tool to access dense nuclear matter.

This approach has been applied to various properties of baryonic matter up to the density commensurate with normal nuclear matter, making connection with Landau Fermi-liquid fixed point theory. The application relies on Wilsonian renormalization-group strategies~\cite{BR:DD}. Early discussions on this matter can be found in \cite{chaejun}.

The novelty in applying the sliding-vacuum strategy to density regimes higher than $n_0$ is that the intrinsic density dependence of QCD condensates in the parameters of the Lagrangian  takes into account the topology change observed in the skyrmion crystal model. The topology change that takes place at $n_{1/2}\sim 2n_0$ is then encoded in how the various parameters of the Lagrangian behaves as density goes from below to above $n_{1/2}$.

The characteristic features resulting from $bs$HLS with the topology change at $n=n_{1/2}$ incorporated are: As density goes higher than $n_{1/2}$, (1) the baryonic matter flows to the vector manifestation at which the $\rho$ mass $m_\rho^\ast \to 0$~\cite{Harada:2000kb},  (2) the nucleon mass $m_N^\ast\to f_\pi^\ast\approx f_\sigma^\ast\approx m_0\neq 0$ and (3) such that  $m_\rho^\ast/m_N^\ast\to 0$.  With these taken into account and with $n_{1/2}\approx 2n_0$,  the theory has been applied to both normal nuclear matter and dense compact-star matter using the well-established $V_{lowk}$ {renormalization group (RG)} technique~\cite{Dong:2012ch,PKLR}. This approach is verified to work well for nuclear properties and predicts the mass and radius in fair agreement with the observed massive stars~\cite{Demorest:2010bx,Antoniadis:2013pzd}.

\begin{figure}[H]\centering
\includegraphics[scale=0.3,angle=270]{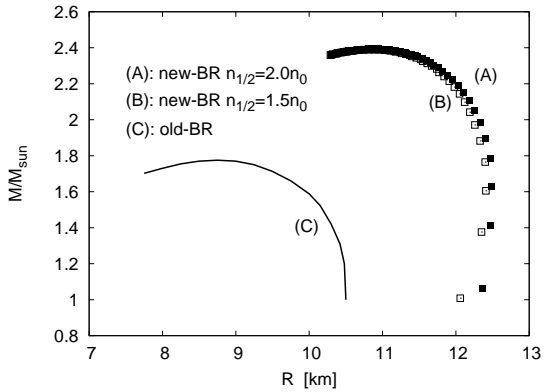}
\caption{Mass-radius relation calculated from the $bs$HLS~\cite{Dong:2012ch}. (C) is gotten with no topology change and (A) \& (B) with topology change.}
\label{fig:MvsRBRSnHf}
\end{figure}

A full description of the calculation is rather involved, including the subtle issue of double decimations in doing RG calculation in nuclear systems, etc. It suffices for our purpose to give a few results to illustrate the main point of this subsection.
\begin{figure}[H]\centering
%\begin{center}
\includegraphics[height=5.9cm]{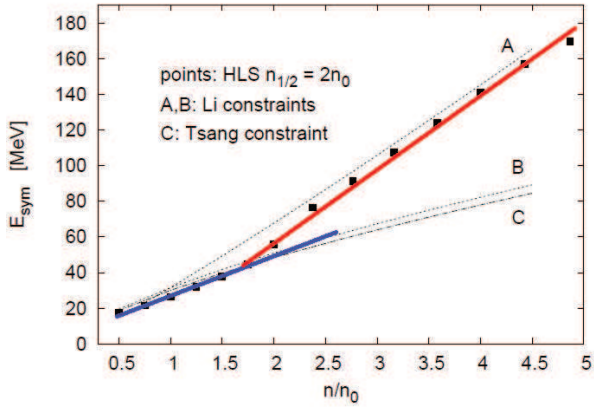}
\caption{The symmetry energy predicted by the $V_{lowk}$ with $bs$HLS with $n_{1/2}=2n_0$~\cite{Dong:2012ch}. The eye-ball slope change is indicated by the colored straight lines. The lines labelled as A, B, C are empirical constraints coming from heavy-ion experiments.}\label{Esym}
%\label{tensor}
%\end{center}
\end{figure}
First of all, it gives a unique prediction for the symmetry energy Fig.~\ref{Esym},  weeding out the wilderness in Fig.~\ref{fig:SymmE}. The topology change that gave the cusp in the skyrmion crystal is manifested here as a changeover from soft-to-hard EoS at $n_{1/2}$. Secondly, as shown in Fig.~\ref{fig:MvsRBRSnHf}, the topology change increases dramatically the maximum mass from (C) $\sim 1.7$ M$_\odot$ (no topology change) to (A) or (B) $ \gsim 2$ M$_\odot$ (topology change). The changeover resembles hadron-quarkyonic transition~\cite{fukushima}, suggesting a quark-hadron duality.

\begin{figure}[H]
\begin{center}
\includegraphics[height=5.9cm]{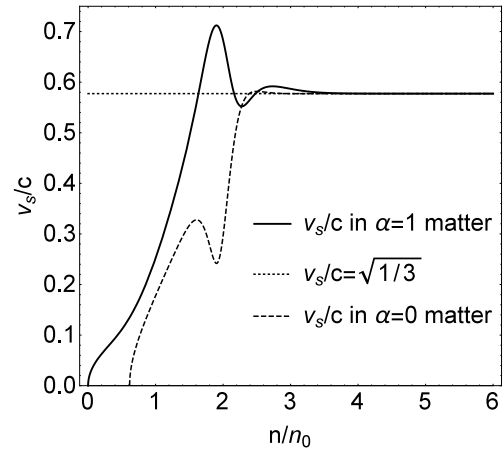}
\caption{
The sound velocity predicted in $bs$HLS model~\cite{PR-emergence}. $\alpha\equiv (N-Z)/(N+Z)$ where $N(Z)$ is the neutron(proton) number.}
\label{sound}
\end{center}
\end{figure}

Another feature, hitherto undiscovered in the field, is the possibility that the sound velocity of massive compact stars could approach, for $n> n_{1/2}\sim 2n_0$, what is conventionally referred to as ``conformal," $v_s^2/c^2=1/3$~\cite{PR-emergence}. This is shown in Fig.~\ref{sound}. The trace of energy momentum tensor is not equal to zero but density-independent constant, so the half-skyrmion state is not strictly scale-invariant but hints at the dilaton limit fixed point at which the dilaton condensate goes toward zero. This is an extremely interesting possibiltiy to explore experimentally.

\section{Perspectives and Discussions}

\label{sec:dis}

Skyrme's pioneering idea of getting baryons from  mesonic theories was not only to unify the mesons and baryons in a single framework but also to treat the nucleon, elementary particle, and the nucleus, complex system of many nucleons, on the same footing. One is then to arrive at nucleus from pion as one does to nucleon. It is now established that the skyrmion out of the pion field is a nucleon in the large $N_c$ limit of QCD in which quarks and gluons are the elementary constituents. In making these multiple connections, at the core lies the  topology. In fact, topology figures strikingly in all aspects of physics, as one can gather from the volume entitled ``The Multifaceted Skyrmion"~\cite{BReditor}, ranging from quarks/gluons to nucleons to nuclei to condensed matter to string theory. Numerous startling new discoveries are being made involving topology in 2 and 3 dimensional systems, such as  topological supeconductivity, and some daring mathematical framework is brought out to organize all visible matters, molecules, atoms, nuclei, in complex geometry~\cite{atiyah-manton}. While a great progress with amazing applications is being made in condensed matter, with new experimental discoveries, the progress in nuclear physics for which Skyrme's original idea was put forward has met with much less success. This has in part to do with that direct experimental observations are difficult to come by and theories, involving strong interactions, are harder to control.

There, however, have been efforts since a decade or so, with some notable progress in difficult fundamental problems of nuclear physics. The principal effort has been directed to decipher what takes place at high density relevant to the interior of compact stars, an extremely difficult problem which is currently out of reach by the only known nonperturbative tool of QCD, lattice QCD, at high density.

The progress made so far comes from several different directions. With the understanding of the Cheshire Cat Phenomenon, what is confined -- color -- and what is not confined -- baryon charge -- is clarified in terms of quark-bag-skyrmion duality.  This then hints at the origin of the proton mass via an intricate interplay involving topology between confinement, a mathematically daunting problem, and spontaneous breaking of chiral symmetry~\cite{MR:mass}. It is now becoming clear that the origin of mass for bosons is most likely different from that for baryons.  It is the topology change from skyrmions to half-skyrmions in dense matter at $n\gsim 2n_0$ that gives the signal for this difference. In the density regime $n\gsim n_{1/2}\sim 2n_0$, the hidden gauge symmetry associated with the vector mesons and the hidden scale symmetry associated with the dilaton scalar intervene in nuclear dynamics in such a way that  while the $\rho$ mass tends toward zero, the nucleon mass goes to a chirally invariant mass $m_0$ which is non-zero and large $m_0\sim O(m_N)$. Thus going toward possible deconfinement density $n_{dc} \gg n_0$, the ratio of the effective masses in medium $m_\rho/m_N\to 0$, which is at odds with the standard ``Nambu scenario" for the origin of mass based on the NJL model.

In the range of densities relevant to compact stars, there is neither indication nor need for quark-gluon degrees of freedom. In fact, the EoS described by the topology change at $n_{1/2}\sim 2n_0$ has a certain similarity to the smooth hadron-quark or hadron-quarkyonic phase change that is assumed to take place at $\sim (2-3)n_0$~\cite{fukushima}. It is also possible that the half-skyrmion phase -- which may be dual to the quarkyonic phase -- is a precursor to (emergent) scale invariance with the sound velocity $v_s^2/c^2=1/3$. In the spirit of the Cheshire Cat, this seems to suggest that ``deconfinement" is irrelevant in compact-star physics (See \cite{glozman} for an argument that there is no deconfinement in QCD.). If deconfinement should intervene at some density, this would mean that the treatment anchored on topology should break down at that density. Clarifying this issue is beyond the framework that is taken in the works reviewed here.

%%%%%%%%%%%%%%%%%%%%%%%%%%%%%%%%%%%%%%%%%%%%%%%%%%%%%%%%%%%%%%%%%%%%%%%%%%%%%%%%%%%%%%%%%%%%%%%%%%

\vspace*{2mm} \Acknowledgements{\bahao The work of Y.-L.~M. was supported in part by National Science Foundation of China (NSFC) under
Grant No.~11475071, 11547308 and the Seeds Funding of Jilin University.}

\end{multicols}

\end{document}